\documentclass[]{article}
\usepackage{lmodern}
\usepackage{amssymb,amsmath}
\usepackage{ifxetex,ifluatex}
\usepackage{fixltx2e} 
\ifnum 0\ifxetex 1\fi\ifluatex 1\fi=0 
  \usepackage[T1]{fontenc}
  \usepackage[utf8]{inputenc}
\else 
  \ifxetex
    \usepackage{mathspec}
  \else
    \usepackage{fontspec}
  \fi
  \defaultfontfeatures{Ligatures=TeX,Scale=MatchLowercase}
\fi
\IfFileExists{upquote.sty}{\usepackage{upquote}}{}
\IfFileExists{microtype.sty}{%
\usepackage{microtype}
\UseMicrotypeSet[protrusion]{basicmath} 
}{}
\usepackage[margin=1in]{geometry}
\usepackage{hyperref}
\hypersetup{unicode=true,
            pdftitle={Spillover Effects in Experimental Data},
            pdfauthor={Peter M. Aronow, Yale; Dean Eckles, MIT; Cyrus Samii, NYU; Stephanie Zonszein, NYU},
            pdfborder={0 0 0},
            breaklinks=true}
\usepackage{natbib}
\usepackage{color}
\usepackage{fancyvrb}

\DefineVerbatimEnvironment{Highlighting}{Verbatim}{commandchars=\\\{\}}
\usepackage{framed}
\definecolor{shadecolor}{RGB}{248,248,248}
\newenvironment{Shaded}{\begin{snugshade}}{\end{snugshade}}
\newcommand{\KeywordTok}[1]{\textcolor[rgb]{0.13,0.29,0.53}{\textbf{#1}}}
\newcommand{\DataTypeTok}[1]{\textcolor[rgb]{0.13,0.29,0.53}{#1}}
\newcommand{\DecValTok}[1]{\textcolor[rgb]{0.00,0.00,0.81}{#1}}

\newcommand{\FloatTok}[1]{\textcolor[rgb]{0.00,0.00,0.81}{#1}}

\newcommand{\StringTok}[1]{\textcolor[rgb]{0.31,0.60,0.02}{#1}}

\newcommand{\OtherTok}[1]{\textcolor[rgb]{0.56,0.35,0.01}{#1}}

\newcommand{\OperatorTok}[1]{\textcolor[rgb]{0.81,0.36,0.00}{\textbf{#1}}}

\newcommand{\NormalTok}[1]{#1}
\usepackage{graphicx,grffile}
\makeatletter
\def\maxwidth{\ifdim\Gin@nat@width>\linewidth\linewidth\else\Gin@nat@width\fi}
\def\maxheight{\ifdim\Gin@nat@height>\textheight\textheight\else\Gin@nat@height\fi}
\makeatother
\setkeys{Gin}{width=\maxwidth,height=\maxheight,keepaspectratio}
\ifx\paragraph\undefined\else
\let\oldparagraph\paragraph
\renewcommand{\paragraph}[1]{\oldparagraph{#1}\mbox{}}
\fi
\ifx\subparagraph\undefined\else
\let\oldsubparagraph\subparagraph
\renewcommand{\subparagraph}[1]{\oldsubparagraph{#1}\mbox{}}
\fi

\let\rmarkdownfootnote\footnote%
\def\footnote{\protect\rmarkdownfootnote}

\usepackage{titling}

  \title{Spillover Effects in Experimental Data}
    \pretitle{\vspace{\droptitle}\centering\huge}
  \posttitle{\par}
    \author{Peter M. Aronow, Yale \\ Dean Eckles, MIT \\ Cyrus Samii, NYU \\ Stephanie Zonszein, NYU}
    \preauthor{\centering\large\emph}
  \postauthor{\par}
    \date{}
    \predate{}\postdate{}
  
\usepackage{booktabs}
\usepackage{longtable}
\usepackage{array}
\usepackage{multirow}
\usepackage{wrapfig}
\usepackage{float}
\usepackage{colortbl}
\usepackage{pdflscape}
\usepackage{tabu}
\usepackage{threeparttable}
\usepackage{threeparttablex}
\usepackage[normalem]{ulem}
\usepackage{makecell}
\usepackage{xcolor}

\usepackage{setspace}

\begin{document}
\maketitle
\begin{abstract}
We present current methods for estimating treatment effects and
spillover effects under ``interference'', a term which covers a broad
class of situations in which a unit's outcome depends not only on
treatments received by that unit, but also on treatments received by
other units. To the extent that units react to each other, interact, or
otherwise transmit effects of treatments, valid inference requires that
we account for such interference, which is a departure from the
traditional assumption that units' outcomes are affected only by their
own treatment assignment. Interference and associated spillovers may be
a nuisance or they may be of substantive interest to the researcher. In
this chapter, we focus on interference in the context of randomized
experiments. We review methods for when interference happens in a
general network setting. We then consider the special case where
interference is contained within a hierarchical structure. Finally, we
discuss the relationship between interference and contagion. We use the
\emph{interference} R package and simulated data to illustrate key
points. We consider efficient designs that allow for estimation of the
treatment and spillover effects and discuss recent empirical studies
that try to capture such effects.
\end{abstract}

\newcommand{\X}{\mathbf{X}}
\newcommand{\Y}{\mathbf{Y}}
\newcommand{\Z}{\mathbf{Z}}
\newcommand{\I}{\mathbf{I}}
\newcommand{\V}{\mathbf{V}}
\newcommand{\W}{\mathbf{W}}
\newcommand{\R}{\mathbf{R}}
\newcommand{\Q}{\mathbf{Q}}
\newcommand{\A}{\mathbf{A}}
\newcommand{\B}{\mathbf{B}}
\newcommand{\D}{\mathbf{D}}
\newcommand{\T}{\mathbf{T}}
\newcommand{\F}{\mathbf{F}}
\newcommand{\M}{\mathbf{M}}

\renewcommand{\O}{\mathbf{O}}
\renewcommand{\H}{\mathbf{H}}
\renewcommand{\L}{\mathbf{L}}
\newcommand{\g}{\mathbf{g}}
\newcommand{\tht}{\text{\tiny{HT}}}
\newcommand{\thj}{\text{\tiny{HJ}}}
\newcommand{\ttl}{\text{\tiny{TL}}}
\newcommand{\tts}{\text{\tiny{TS}}}
\newcommand{\tr}{\text{\tiny{R}}}
\newcommand{\td}{\text{\tiny{D}}}
\renewcommand{\d}{\mathbf{d}}
\newcommand{\matp}{\mathbf{p}}
\renewcommand{\r}{\mathbf{r}}
\newcommand{\di}{\text{di}}
\newcommand{\vdi}{\text{vdi}}
\newcommand{\e}{\mathbf{e}}
\newcommand{\boldt}{\mathbf{t}}
\newcommand{\s}{\mathbf{s}}
\newcommand{\uu}{\mathbf{u}}
\newcommand{\w}{\mathbf{w}}
\newcommand{\x}{\mathbf{x}}
\newcommand{\vv}{\mathbf{v}}
\newcommand{\y}{\mathbf{y}}
\newcommand{\z}{\mathbf{z}}
\newcommand{\ident}{\mathbf{i}}
\newcommand{\matc}{\mathbf{c}}
\newcommand{\m}{\mathbf{m}}
\newcommand{\bt}{\mathbf{t}}
\newcommand{\Var}{\mathrm{Var}}
\newcommand{\Cov}{\mathrm{Cov}}

\newcommand{\Beta}{\boldsymbol{\beta}}
\newcommand{\btheta}{\boldsymbol{\theta}}
\newcommand{\bgamma}{\boldsymbol{\gamma}}
\newcommand{\bpi}{\boldsymbol{\pi}}
\newcommand{\arrowp}{\stackrel{p}{\rightarrow}}
\newcommand{\0}{\mathbf{0}}
\newcommand{\bP}{\mathbf{P}}
\newcommand{\duby}{\mathbf{y}}
\newcommand{\dubx}{\mathbb{x}}
\newcommand{\dubd}{\mathbb{d}}
\newcommand{\dubp}{\mathbb{p}}
\newcommand{\dubi}{\mathbb{i}}
\newcommand{\dubR}{\mathbb{R}}
\newcommand{\dubpi}{\mathbb{\pi}}
\newcommand{\dipi}{\Pi^{-1}}
\newcommand{\dipit}{\Pi^{-1}_1}
\newcommand{\dipic}{\Pi^{-1}_0}
\newcommand{\diR}{\mathbf{R}}
\newcommand{\E}{\mathbb{E}}

\clearpage

\section{Introduction}\label{introduction}

We present current methods for identifying causal effects under
``interference'', a term which covers a very broad class of situations
in which a unit's outcomes depend not only on treatments received by
that unit, but also on treatments received by other units \citep{cox58}.
This includes effects that spill over from one unit to others. For
example, in an agricultural experiment, fertilizer applied to one plot
of land may literally spill over into other plots assigned to different
treatments, therefore affecting their yields. Interference may arise
from interactions between units, such as through social influence
processes. For example, in the context of an election, exposing a voter
to a persuasive appeal may affect what that voter says to their friends,
which in turn may affect the friends' outcomes.

Interference represents a departure from the traditional assumption
wherein the potential outcomes that would be observed for a unit in
either the treatment or control condition, depend only on that unit's,
and not the overall, treatment assignment. This traditional assumption
is implied by what \citet{rubin1990} refers to as the ``stable unit
treatment value assumption'' (SUTVA).

\begin{figure}
\centering
\includegraphics[width=0.60000\textwidth]{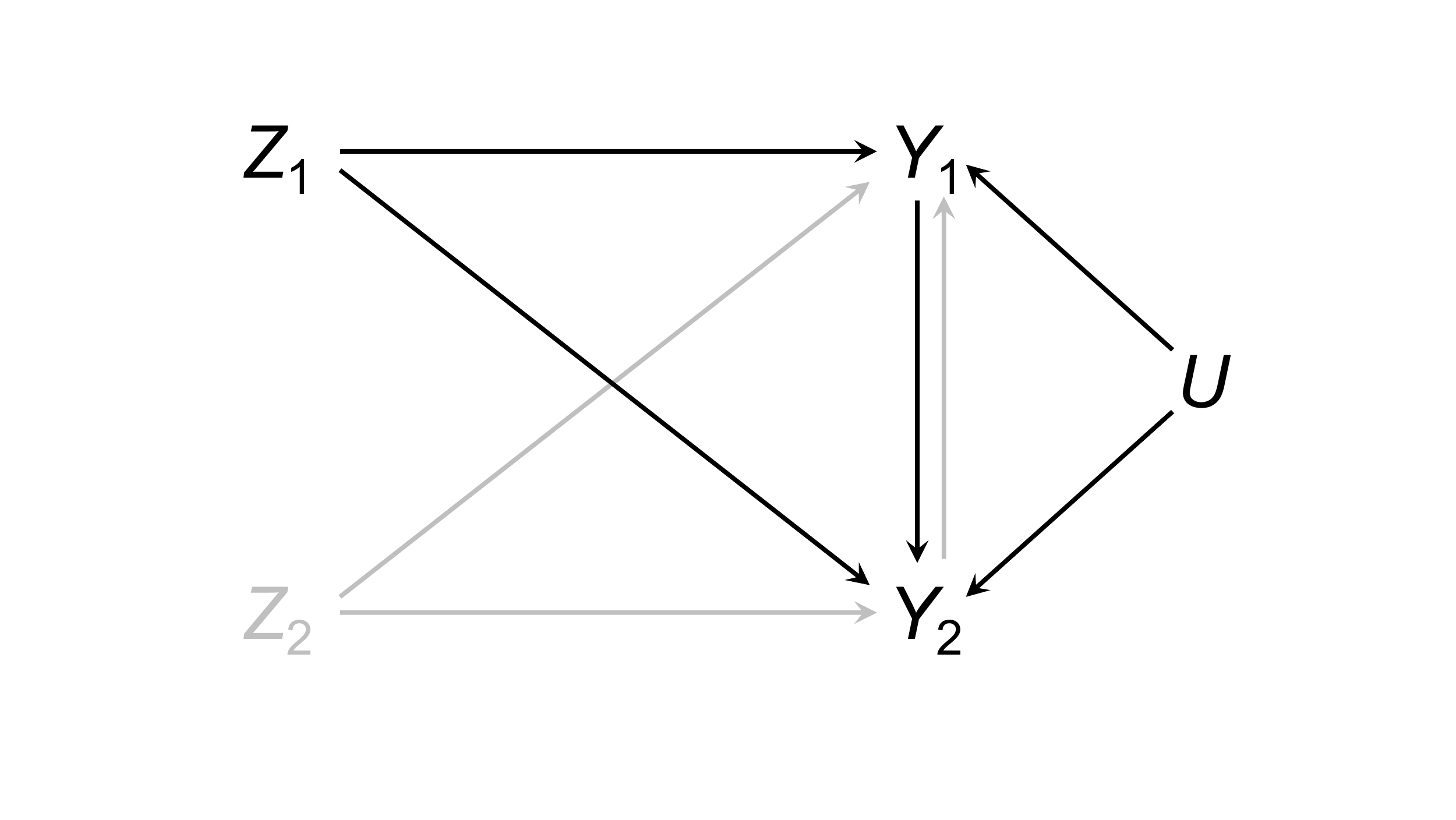}
\caption{Causal graph illustrating interference mechanisms and
confounding mechanisms when treatment (\(Z\)) is randomized.
\label{fig:interference_dag}}
\end{figure}

Figure \ref{fig:interference_dag} displays channels through which
interference might occur. The black elements in Figure
\ref{fig:interference_dag} show a directed acyclic graph that captures
potential spillover effects onto unit 2 from a treatment assigned to
unit 1. We assume an experiment where unit 1's treatment, \(Z_1\), is
randomly assigned. Then, the effect of this treatment, captured by the
black arrows flowing from \(Z_1\), could be to alter unit 1's own
outcome, \(Y_1\), as well as unit 2's outcome, \(Y_2\). This could
happen via a pathway in which \(Y_1\) mediates the effect of \(Z_1\) on
\(Y_2\). Such outcome-mediated effects are known as \emph{contagion}
effects, which we briefly discuss toward the end of this chapter. Or it
could be that \(Z_1\) affects \(Y_2\) through channels that do not go
through \(Y_1\). Spillover includes the sum total of the effects from
\(Z_1\) to \(Y_2\). The gray elements in Figure
\ref{fig:interference_dag} show that spillovers could be running from
\(Z_2\) to \(Y_1\) as well. Finally, the variable \(U\) captures other
variables that might induce dependency between \(Y_1\) and \(Y_2\). It
is important to recognize that these sources of outcome dependence or
clustering are wholly distinct from spillover. However, such confounders
undermine the ability to isolate contagion effects from other spillover
mechanisms, a point to which we return below.

In this chapter, we assume that the researcher is interested in
estimating spillover effects. We focus on randomized experiments for
which we have some understanding of the structure through which
spillover effects occur. In the first section, we review cases where the
interference network is known completely, but then can take almost
arbitrary form. In the second section, we review cases where we know
only that interference is fully contained within the boundaries of
strata that partition the population, but then the interference network
within these strata is unknown. The final section makes some points
regarding the attempt to distinguish contagion effects from other forms
of spillover. We do not discuss work that examine causal effects in
situations where the interference network is fully hidden as in
\citet{savje2017average}. Moreover, our emphasis is on large-N
estimation of spillover effects, and so we omit discussion of methods
that work off the experimental randomization to develop exact tests for
interference effects
\citep{rosenbaum07_interference, aronow2012general, bowers2013reasoning, athey2018exact}.
These methods either restrict themselves to testing for interference, or
require strong assumptions (for example, a finite-dimensional model of
causal effects) to attain interval estimates.

\section{Three Motivating Examples}\label{three-motivating-examples}

We begin with three examples that allow us to illustrate key points. The
first example is a study by \citet{paluck2016changing}, who study the
effects of an antibullying program in New Jersey schools. The authors
began by measuring the schools' social networks. They did this by asking
students to report which other students they chose to spend time with in
previous weeks. The experiment then randomly assigned schools to the
antibullying program, and within schools, randomly selected students
from an eligible subpopulation to actively participate in the program.
Assuming the network measures are accurate, the experiment identifies
spillover effects onto students who themselves do not participate in the
program but have peers who do. The ability to get at such spillover
effects depends on the accuracy of the network measure and the ways that
one specifies potential exposure to spillovers on the basis of this
network. If the researcher assumes that only peers of program recipients
can be affected by the anticonflict intervention, but in fact peers of
peers can be affected as well, then inferences about the program's
direct and spillover effects may be biased. The section below on
arbitrary interference networks discuss this issue along with associated
sensitivity analyses.

\begin{figure}
\centering
\includegraphics[width=0.50000\textwidth]{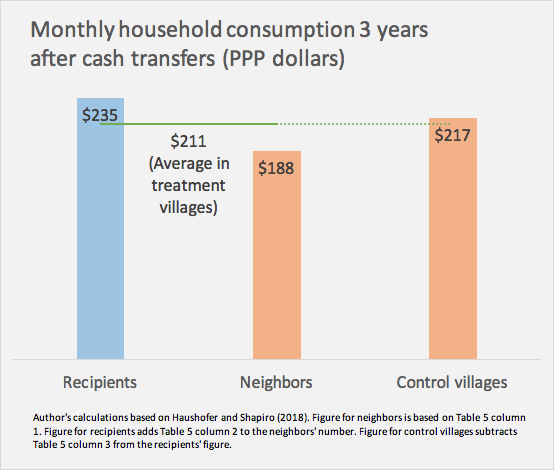}
\caption{Figure from \citet{sandefur2018bloomberg} displaying long-term
spillover effects from an unconditional cash transfer program, as
reported in \citet{haushofer2018long}. \label{fig:negative_spillover}}
\end{figure}

As a second example, consider Figure \ref{fig:negative_spillover}, which
shows the results of an experimental study in Kenya by
\citet{haushofer2018long} on the long-term (after 3 years) effects of
unconditional cash transfers. The outcome here is monthly household
consumption. In this study, villages were randomly selected to be
treated, and then within these villages, a half of the households were
randomly selected to receive about 400 USD in cash. As Figure
\ref{fig:negative_spillover} shows, average monthly consumption in
treatment villages, pooling recipient and non-recipient households, is
\$211 (green horizontal line), which is similar to the average in
control villages (\$217). But the inter-village comparison masks
variation within treated villages. Recipient households consume \$235
per month on average, while non-recipient neighbors in treatment
villages consume \$188 per month on average. Given that the treatment
was assigned in a manner that randomized both across villages and within
villages, these three types of households (recipients, neighbors, and
control villages) are \emph{ex ante} exchangeable, in which case the
experiment yields an unbiased estimate of a negative within-village
spillover effect.

This example illustrates a few additional nuances. First, when
spillovers are present, one needs to think about effects in terms of
overall assignment patterns. Neither the comparison between directly
treated households and control village households nor directly treated
households and untreated households in treated villages gets at ``the''
effect of cash transfers. Effects depend on exactly whom and at what
rate potentially interacting households are treated. This particular
experiment gives evidence on outcomes for treated households and
untreated neighboring households when the within-village treatment rate
is 50\%. Were the quantity of policy interest to be how outcomes change
going from 0\% to 100\% treatment rates, for example, it is not clear
that the experiment could directly speak to that. Second, households
presumably differ not only in how they would respond after receiving the
transfer per se, but also in how they would respond given the precise
set of other households that receive the transfer. Suppose the village
included two business partners, and business production exhibited
increasing returns to input capital. Then, outcomes would presumably
differ if the households of both partners were treated as compared to
only one partner being treated. Outcomes among the untreated neighbors
in treatment villages depend not only on the treatment saturation rate
(50\%), but also the precise configuration of treated and untreated
households. This raises the question of how to interpret effects such as
those presented in Figure \ref{fig:negative_spillover}---what
counterfactual comparisons are being characterized, exactly? The section
below on partial interference addresses these issues.

A third example is the experiment by \citet{nickerson2008voting} on
potential contagion in voter turnout. Households with two registered
voters were first randomly assigned to one of three conditions: a
get-out-the-vote (GOTV) doorstep appeal, a doorstep appeal to promote
recycling, or a control condition in which nothing was done to the
household. The recycling appeal was meant to serve as a ``placebo''
treatment to account for the fact that only some households, and
particularly individuals with specific characteristics would open the
door to receive an appeal. The experiment thus yielded data on subjects
who opened the door and were thus direct recipients of either the GOTV
or recycling appeals, their housemates who were not there at the door,
and then the full set of control households. Comparing voter turnout
among housemates of those who directly received the GOTV appeal to
housemates of those who received the recycling appeal, one can estimate
whether the GOTV treatment spilled over from the direct recipient to the
housemate. Insofar as there is an effect, one may wonder if the
mechanism at work is contagion---that is, it is the voting intention of
the direct recipient that went on to affect the voting intention of the
housemate---or some other mechanism. This distinction between mechanisms
would have implications for theories about norms that support voting
behavior. In the section below on contagion, we return to this
experiment and review assumptions needed to isolate contagion effects.

\section{Formal Setting}\label{formal-setting}

We now present a formal framework for defining causal effects under
interference. Suppose an experimenter intervenes on a finite population
\(U\) of units indexed by \(i = 1, \ldots, N\). Let us suppose further
that the intervention is defined by a treatment assignment vector
\(\z = (z_1, \ldots, z_N)'\), where \(z_i \in \{1,0\}\) specifies the
possible treatment values that unit \(i\) receives. Let \(\Omega\) be
the set of treatment assignment vectors \(\z\) with
\(\left\vert{\Omega}\right\vert = 2^N\). An \emph{experimental design}
is a plan for randomly selecting a particular value of \(\z\) from the
\(2^N\) different possibilities with predetermined probability
\(p_\z\)---for example, Bernoulli assignment (i.e., coin flips) or
completely randomized assignment strategy. Therefore,
\(\Omega = \{\z : p_\z >0 \}\) and the realized treatment assignment
\(\Z = (Z_1, \ldots, Z_N)'\) is a random vector with support \(\Omega\)
and \(\Pr(\Z = \z) = p_\z\). For example, with a population of size
\(N=10\), and an experimental design that randomly assigns without
replacement a proportion \(p=0.2\) to treatment condition \(z_i =1\)
with uniform probability, there are \({N\choose pN}=45\) possible
treatment assignments (\(\left\vert{\Omega}\right\vert = 45\)) and the
realized treatment assignment \(\Z\) has \(p_\z = \frac{1}{45}\). The
experimental design characterizes precisely the probability distribution
of the assigned treatments. In experiments, this is determined by the
researcher and is therefore known.

To analyze the effect of different treatment assignments, we compare the
different outcomes they produce. These potential outcomes are defined
for each unit \(i\) as the elements in the image of a function that maps
assignment vectors to a real valued outcomes,
\(y_i : \Omega \rightarrow \dubR\). Particularly, \(y_i(\z)\) is the
response of unit \(i\) to assignment \(\z\). For convenience, let
\(\z_{-i}=(z_i, \ldots, z_{i-1},z_{i+1}, \ldots, z_N)'\) denote the
\((N-1)\)-element vector that removes the \(i\)th element from \(\z\).
Then, the potential outcome \(y_i(\z)\) can equivalently be expressed as
\(y_i(z_i; \z_{-i})\). Continuing with the example of the cash transfer
program above, this quantity would be the potential consumption of
household \(i\) given its assignment as a transfer recipient or
non-recipient (\(z_i\)) and the treatment assignment of all other
households (\(\z_{-i}\)), including those inside and outside household
\(i\)'s village.

Traditional analyses of experiments, and other chapters in this volume,
assume no interference, in which case the potential outcome \(y_i(\z)\)
is restricted to be affected only by \(i\)'s own treatment. That is,
with no interference, for any two treatment assignments \(\z\) and
\(\z'\), for which \(z_i\) remains unchanged, we have
\(y_i(z_i; \z_{-i}) = y_i(z_i; \z_{-i}')\) for all
\(i \in \{1, \ldots, N\}\). When interference is present, there exist
some units \(i \in U\) for which
\(\ y_i(z_i; \z_{-i}) \neq y_i(z_i; \z_{-i}')\), that is fixing the
treatment of \(i\) while changing other units' treatment results in
changes of \(i\)'s outcome.

Let \(Y_i\) denote the observed outcome of unit \(i\), where the
observed outcome is related to the potential outcomes as
\(Y_i = y_i(\Z) = y_i(Z_i;\Z_{-i})\), where \(\Z_{-i}\) denotes the
vector \(\Z\) net of its \(i\)th element. In the case of no
interference, \(Y_i = y_i(Z_i)\). Therefore, when interference is
present, we need to account for others' treatment assignments as well.

\section{Arbitrary But Known Interference
Networks}\label{arbitrary-but-known-interference-networks}

This section reviews estimation methods in a setting where interference
occurs over a network of arbitrary structure, but this structure is
known. The analysis follows \citet{aronow2017estimating}. We represent a
unit's set of interfering units in terms of network ties. Then,
depending on the network structure and the treated units' network
characteristics, different treatment assignments may result in different
and arbitrary, but known, patterns of interference. For example,
assuming that interference happens through direct ties between units,
treating any one unit in a fully-connected network generates a pattern
in which the treatment of that one unit interferes with the treatment of
every other unit in the network. In a regular lattice, the treatment of
any one treated unit interferes only with the treatment of that unit's
four nearest neighbors, and in an irregular network, treatment
assignments that treat units with many direct ties generate more
interference, than assignments that treat units with just a few ties.

As in the anticonflict social network experiment of
\citet{paluck2016changing}, these methods require the researcher to
measure the network or to have comprehensive information about
connections between experimental units, and to define precise causal
effects which reflect the possible types of treatment \emph{exposures}
that might be induced in the experiment, which in turn requires to make
specific assumptions about the extent of interference. The goal is to
estimate exposure-specific causal effects---for the anticonflict
program, for example, we might estimate effects on students for whom at
least one peer is a direct program participant, or for whom exactly two
peers are participants, etc. Knowing the treatment assignment
distribution allows one to account for potential sources of confounding
that arise from heterogeneity across units in their likelihood of
falling into different exposure conditions (for example, heterogeneity
in terms of students' number of connections with other students). The
sections below explain.

\subsection{Exposure Mapping}\label{exposure-mapping}

To determine each unit's treatment exposure under a given treatment
assignment, \citet{aronow2017estimating} define an \emph{exposure
mapping} that maps the set of assignment vectors and unit-specific
traits to an exposure value:
\(f : \Omega \times \Theta \rightarrow \Delta\), where
\(\theta_i \in \Theta\) quantifies relevant traits of unit \(i\) such as
the number of direct ties to other units in the network and, possibly,
weights assigned to each of these ties. The set \(\Delta\) contains all
of the possible treatment-induced exposures that may be generated in the
experiment, and its cardinality depends on the nature of interference.
For example, with no interference and a binary treatment the exposure
mapping ignores unit specific traits \(f(\z, \theta_i) = z_i\),
producing two possible exposure values for each unit: no exposure (or
control condition, \(z_i=0\)) and direct exposure (or treatment
condition, \(z_i=1\)), in which case \(\Delta = \{0, 1\}\). Now,
consider interference that occurs through direct peer connections. Then,
\(\theta_i\) is a column vector equal to the transpose of unit \(i\)'s
row in a network adjacency matrix (which captures \(i\)'s direct
connections to other units), and the exposure mapping
\(f(\z, \theta_i)\) can be simply defined to capture \emph{direct}
exposure to treatment---or the effect of being assigned to
treatment---and \emph{indirect} exposure---or the effect of being
exposed to treatment of peers.\footnote{Note that the meaning of
  ``direct'' and ``indirect'' in the interference setting is different
  than in the mediation setting reviewed in Glynn's chapter in this
  volume.} An example of such an exposure mapping (and by no means the
only possibility) is the following, whereby indirect exposure occurs
when at least one peer is treated:

\[f(\z, \theta_i) = \begin{cases} d_{11} (\text{Direct + Indirect Exposure}): & z_i\I(\z'\theta_i>0) =1, \\
d_{10} (\text{Isolated Direct Exposure}): & z_i\I(\z'\theta_i=0) =1, \\
d_{01} (\text{Indirect Exposure}): & (1-z_i)\I(\z'\theta_i>0) =1,\\
d_{00} (\text{No Exposure}): & (1-z_i)\I(\z'\theta_i=0) =1 \end{cases}\]

\noindent For this particular case
\(\Delta = \{d_{11}, d_{10}, d_{01}, d_{00}\}\). This characterization
of exposures is ``reduced form'' in that it does not distinguish between
the mechanisms through which spillover effects occur.

Specification of the exposure mapping requires substantive consideration
of the data generating process. \citet{manski2013identification}
discusses subtleties that arise in specifying exposure mappings. For
example, the author shows how models of simultaneous endogenous choice
(due to homophily or common external shocks) can produce restrictions on
the potential outcomes \(y_i(d_k)\), and therefore imply that potential
outcomes may vary in ways that an otherwise intuitive exposure mapping
may fail to capture.

Because units occupy different positions in the interference network,
their probabilities of being in one or another exposure condition vary,
even if treatment is randomly assigned. Insofar as network position also
affects outcomes, then such differences in exposure probabilities need
to be taken into account when estimating exposure-specific causal
effects. Otherwise, the analysis would be confounded. We show here that
when the random assignment mechanism is known, then these exposure
probabilities are also known. This allows one to condition on the
exposure probabilities directly. To see this, define the exposure that
unit \(i\) receives as \(D_i = f(\Z, \theta_i)\), a random variable with
support \(\Delta_i \subseteq \Delta\) and for which
\(\Pr(D_i = d) = \pi_i(d)\). For each unit \(i\) there is a vector,
\(\bpi_i = (\pi_i(d_1), \ldots, \pi_i(d_K))'\), with the probability of
\(i\) being subject to each of the possible exposures in
\(\{d_1, \ldots, d_K\}\). \citet{aronow2017estimating} call \(\bpi_i\)
\(i\)'s \emph{generalized probability of exposure}. For example, the
exposure mapping defined above gives rise to
\(\bpi_i = (\pi_i(d_{11}), \pi_i(d_{10}), \pi_i(d_{01}), \pi_i(d_{00}))'\).
We observe the unit-specific traits (\(\theta_i\)) necessary to define
exposures for any treatment assignment vector, and the probability of
each possible treatment assignment vector (\(p_{\z}\)) is known. This
allows us to compute \(\pi_i(d_k)\) as the expected proportion of
treatment assignments which induce exposure \(d_k\) for unit \(i\). When
the set of possible treatment assignment vectors \(\Omega\) is small,
this can be computed exactly. When \(\Omega\) is large, one can
approximate the \(\pi_i(d_k)\) values with arbitrary precision by taking
a large number of random draws from \(\Omega\).
\citet{aronow2017estimating} discuss considerations for how many draws
are needed so as to keep biases small. This Monte Carlo method may in
some cases require a prohibitive number of draws (for example, if
\(|\Delta|\) is large), but for some specific designs and exposure
mappings it may be possible to compute the \(\pi_i(d_k)\) values via a
dynamic program, as in \citet{ugander2013graph}.

The following toy example illustrates how to compute the exposure
received by each unit and the generalized probability of exposure using
the \texttt{interference} package for R \citep{zonszein-interference}.
Suppose we have a set of \(N=10\) units, we randomly assign (without
replacement) a proportion \(p=0.2\) to treatment condition \(z_i=1\)
with uniform probability. In this case, the realized treatment
assignment \(\Z'\) shows that units 6 and 9 are directly treated.

\begin{Shaded}
\begin{Highlighting}[]
\NormalTok{N <-}\StringTok{ }\DecValTok{10}
\NormalTok{p <-}\StringTok{ }\FloatTok{0.2}
\NormalTok{Z <-}\StringTok{ }\KeywordTok{make_tr_vec_permutation}\NormalTok{(N, p, }\DataTypeTok{R =} \DecValTok{1}\NormalTok{, }\DataTypeTok{seed =} \DecValTok{56}\NormalTok{)}
\NormalTok{Z}
\end{Highlighting}
\end{Shaded}

\begin{verbatim}
     [,1] [,2] [,3] [,4] [,5] [,6] [,7] [,8] [,9] [,10]
[1,]    0    0    0    0    0    1    0    0    1     0
\end{verbatim}

\begin{figure}

{\centering \includegraphics{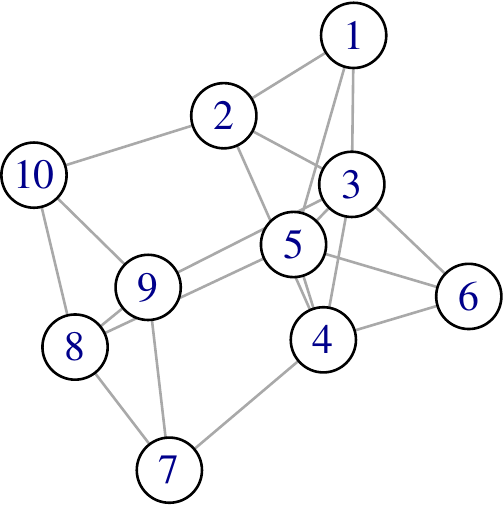} 

}

\caption{\label{fig:small-network} Example of interference network with ten units. Each edge (link) represents a possible channel through which spillover effects might transmit.}\label{fig:unnamed-chunk-4}
\end{figure}

Now let's suppose that units are connected according to a draw from a
Watts-Strogatz model---a random graph generation model that produces
networks with ``small world'' properties: high clustering in network
interconnections and short average path lengths that connect any two
arbitrary nodes (units). We assume an undirected network, that is, if
\(i\) has a direct connection with \(j\), then \(j\) has one with \(i\),
and that on average each unit is directly connected to four other units.
A visualization of such a network is given in Figure
\ref{fig:small-network}, and its adjacency matrix is:

\begin{Shaded}
\begin{Highlighting}[]
\NormalTok{adj_matrix <-}\StringTok{ }\KeywordTok{make_adj_matrix}\NormalTok{(N, }\DataTypeTok{model =} \StringTok{'small_world'}\NormalTok{, }\DataTypeTok{seed =} \DecValTok{492}\NormalTok{)}
\NormalTok{adj_matrix}
\end{Highlighting}
\end{Shaded}

\begin{verbatim}
      [,1] [,2] [,3] [,4] [,5] [,6] [,7] [,8] [,9] [,10]
 [1,]    0    1    1    0    1    0    0    0    0     0
 [2,]    1    0    1    1    0    0    0    0    0     1
 [3,]    1    1    0    1    1    1    0    0    1     0
 [4,]    0    1    1    0    1    1    1    0    0     0
 [5,]    1    0    1    1    0    1    0    1    0     0
 [6,]    0    0    1    1    1    0    0    0    0     0
 [7,]    0    0    0    1    0    0    0    1    1     0
 [8,]    0    0    0    0    1    0    1    0    1     1
 [9,]    0    0    1    0    0    0    1    1    0     1
[10,]    0    1    0    0    0    0    0    1    1     0
\end{verbatim}

For the purposes of our example, the adjacency matrix captures
\(\Theta\), while each row defines \(\theta_i'\). The reason is that in
our example, exposures are defined strictly through the combination of
the treatment assignment and the individual rows of the adjacency
matrix. In principle, exposure mappings could take other factors into
account, such as covariates that are not related to the adjacency matrix
or other properties of the adjacency matrix besides a unit's row.
Returning to the example, from \(\theta_6'\) we know that unit 6 has
edges to each of units 3, 4 and 5. Using the adjacency matrix
(\texttt{adj\_matrix}) and \(\Z\) (\texttt{Z}) as arguments in the
exposure mapping function defined above, we obtain the received exposure
for every unit. The argument \texttt{hop\ =\ 1} describes a data
generating process in which indirect exposure happens through the
existence of any direct peer receiving treatment:

\begin{Shaded}
\begin{Highlighting}[]
\NormalTok{D <-}\StringTok{ }\KeywordTok{make_exposure_map_AS}\NormalTok{(adj_matrix, Z, }\DataTypeTok{hop =} \DecValTok{1}\NormalTok{)}
\NormalTok{D}
\end{Highlighting}
\end{Shaded}

\begin{verbatim}
      dir_ind1 isol_dir ind1 no
 [1,]        0        0    0  1
 [2,]        0        0    0  1
 [3,]        0        0    1  0
 [4,]        0        0    1  0
 [5,]        0        0    1  0
 [6,]        0        1    0  0
 [7,]        0        0    1  0
 [8,]        0        0    1  0
 [9,]        0        1    0  0
[10,]        0        0    1  0
\end{verbatim}

We can see that the received exposure for units 3, 4 and 5 is
\(d_{01} \equiv \text{Indirect Exposure}\), given their direct
connection to unit 6, who is directly treated. Likewise for units 7, 8
and 10 who are degree-one neighbors of unit 9. We can also see that
there are no units in exposure condition \(d_{11}\), because the two
directly treated units (6 and 9) are not connected to each other.

Now, to obtain the generalized probability of exposure of each unit we
need the exposure mapping function and its arguments: the adjacency
matrix and the set of all possible treatment assignments. When
\(\left\vert{\Omega}\right\vert\) is large we can approximate \(\Omega\)
producing random replicate \(\z\)'s. In this case we could easily
compute \(\Omega\) because there are only 45 possible treatment
assignment profiles, but for expository purposes, we work with 30 random
draws from \(\Omega\) without replacement (setting the arguments
\texttt{R\ =\ 30} and \texttt{allow\_repetitions\ =\ FALSE}).

\begin{Shaded}
\begin{Highlighting}[]
\NormalTok{omega <-}\StringTok{ }\KeywordTok{make_tr_vec_permutation}\NormalTok{(}
\NormalTok{  N, p,}
  \DataTypeTok{R =} \DecValTok{30}\NormalTok{, }\DataTypeTok{seed =} \DecValTok{420}\NormalTok{, }\DataTypeTok{allow_repetitions =} \OtherTok{FALSE}
\NormalTok{  )}
\NormalTok{prob_exposure <-}\StringTok{ }\KeywordTok{make_exposure_prob}\NormalTok{(}
\NormalTok{  omega,}
\NormalTok{  adj_matrix,}
\NormalTok{  make_exposure_map_AS,}
  \KeywordTok{list}\NormalTok{(}\DataTypeTok{hop =} \DecValTok{1}\NormalTok{)}
\NormalTok{  )}
\KeywordTok{make_prob_exposure_cond}\NormalTok{(prob_exposure)}
\end{Highlighting}
\end{Shaded}

\begin{verbatim}
               [,1]       [,2]      [,3]       [,4]       [,5]       [,6]
dir_ind1 0.09677419 0.12903226 0.1612903 0.09677419 0.12903226 0.09677419
isol_dir 0.19354839 0.09677419 0.1290323 0.12903226 0.06451613 0.12903226
ind1     0.41935484 0.61290323 0.6774194 0.70967742 0.67741935 0.45161290
no       0.38709677 0.25806452 0.1290323 0.16129032 0.22580645 0.41935484
              [,7]       [,8]      [,9]     [,10]
dir_ind1 0.1290323 0.12903226 0.1290323 0.1290323
isol_dir 0.2258065 0.09677419 0.1612903 0.1290323
ind1     0.4193548 0.64516129 0.6129032 0.4516129
no       0.3225806 0.22580645 0.1935484 0.3870968
\end{verbatim}

The columns capture \(\bpi_i\), the expected proportion of treatment
assignments which result in each exposure. We can see that the exposure
probabilities are very heterogeneous. In this particular case, because
the exposure mapping only takes into account whether a unit is treated
itself and then has at least one treated peer, the exposure
probabilities have a close relationship to the number of direct
connections with peers---i.e., a unit's network degree. Exposure
mappings could be more complex, requiring that one account for different
traits of units (connections to treated peers of peers and other
covariates such as gender or age, for example), in which case exposure
effects would not necessarily map so straightforwardly to network
degree.

\subsection{Spillover Effects as Contrasts Across
Exposures}\label{spillover-effects-as-contrasts-across-exposures}

We now formally define spillover effects as contrasts between averages
of individual potential outcomes across different exposures. To estimate
exposure-specific average potential outcomes, the exposure mapping has
to fully characterize interference. This condition implies that \(K\)
treatment exposures give rise to at most \(K\) distinct potential
outcomes for each unit \(i\) in the population. We write the potential
outcomes as \((y_i(d_1), \dots ,y_i(d_K))\), where
\(y_i(d_k) = y_i(\z)\) for all units, \(k \in \{1, \dots ,K\}\), and
\(\z \in \Omega\) such that \(f(\z, \theta_i)=d_k\). Then, observed
outcomes must relate back to the potential outcomes:
\(Y_i = \sum_{k = 1}^{K} \I(D_i = d_k)y_i(d_k)\). The average potential
outcome at any exposure level \(k\) is then
\(\mu(d_k) = \frac{1}{N} \sum_{i = 1}^{N} y_i(d_k)\), and the average
causal effect of being in exposure condition \(d_k\) as opposed to
exposure condition \(d_l\) is

\[\tau(d_k, d_l) = \frac{1}{N} \sum_{i = 1}^{N} y_i(d_k) - \frac{1}{N} \sum_{i = 1}^{N} y_i(d_l) = \mu(d_k) - \mu(d_l).\]

To estimate \(\sum_{i = 1}^{N} y_i(d_k) = y^T(d_k)\) we have to take
into account that we observe \(y_i(d_k)\) only for those with
\(D_i=d_k\), and that the probability of observing \(y_i(d_k)\) is not
equal across units. Using the exposure mapping from the example above,
the probability of observing
\(y_i(d_{10}) = y_i(\text{Isolated Direct Exposure})\) is smaller for
those with more direct connections to other units in the network. But as
we saw above, by design, we can calculate the probability of the
exposure conditions for each individual. Then, assuming that all units
have nonzero probabilities of being subject to each of the K
exposures\footnote{If \(\pi_i(d_k)=0\) for some units, then estimation
  of average potential outcomes \(\mu(d_k)\) must be restricted to the
  subset of units for which \(\pi_i(d_k)>0\). Interpretation of
  contrasts of average potential outcomes as causal effects would
  require doing so for units such that \(\pi_i(d)>0\) for all
  \(d \in \Delta' \subset \Delta\); for example, when estimating
  \(\tau(d_k, d_l)\) one would need to restrict analysis to units with
  both \(\pi_i(d_k)>0\) and \(\pi_i(d_l)>0\).}, \(y^T(d_k)\) can be
estimated without bias with the Horvitz-Thompson inverse probability
estimator:

\[\widehat{y^T_{HT}}(d_k) = \sum_{i = 1}^{N} \I(D_i = d_k) \frac{Y_i}{\pi_i(d_k)}.\]
In cases where \(\left\vert{\Omega}\right\vert\) is high, and therefore
we use sampling from \(\Omega\) to obtain estimates
\(\hat \pi_i(\cdot)\), those estimates are used in place of the true
\(\pi_i(\cdot)\) values. A Horvitz-Thompson estimator of the average
unit-level causal effect of exposure \(k\) versus \(l\),
\(\tau(d_k, d_l)\), is therefore
\[\widehat{\tau_{HT}}(d_k, d_l) = \widehat{\mu_{HT}}(d_k) - \widehat{\mu_{HT}}(d_l) = \frac{1}{N} \left[\widehat{y^T_{HT}}(d_k) - \widehat{y^T_{HT}}(d_l) \right].\]

Continuing with the previous example, we show how to compute this
estimator. We do so using simulated potential outcomes that exhibit
effect heterogeneity and that vary in units' network degree, in which
case naive estimates that do not account for probabilities of exposure
would be biased. Specifically, we generate a variable with random values
from an absolute standard normal distribution which is correlated with
the unit's first and second order degree---the number of peers and peers
of peers, respectively (for which we use the arguments
\texttt{adj\_matrix} and \texttt{make\_corr\_out} in the function
below). This variable determines the potential outcome under the
\(d_{00}\) condition. To build heterogeneous effects into the analysis,
we assume what \citet{rosenbaum99-dilated} refers to as ``dilated
effects'' such that \(y_i(d_{11})=2 \times y_i(d_{00})\),
\(y_i(d_{10})=1.5 \times y_i(d_{00})\),
\(y_i(d_{01})=1.25 \times y_i(d_{00})\). (The multipliers of
\(y_i(d_{00})\) can be changed by passing a vector with 3 numbers to
\texttt{multipliers} in the function below).

\begin{Shaded}
\begin{Highlighting}[]
\NormalTok{potential_outcomes <-}\StringTok{ }\KeywordTok{make_dilated_out}\NormalTok{(}
\NormalTok{  adj_matrix, make_corr_out, }\DataTypeTok{seed =} \DecValTok{1101}\NormalTok{,}
   \DataTypeTok{multipliers =} \OtherTok{NULL}\NormalTok{, }\DataTypeTok{hop =} \DecValTok{1}
\NormalTok{  )}
\end{Highlighting}
\end{Shaded}

From the potential outcomes and received exposures (\texttt{D}), we get
the observed outcomes.

\begin{Shaded}
\begin{Highlighting}[]
\NormalTok{observed_outcomes <-}\StringTok{ }\KeywordTok{rowSums}\NormalTok{(D}\OperatorTok{*}\KeywordTok{t}\NormalTok{(potential_outcomes))}
\end{Highlighting}
\end{Shaded}

Next, we compute \(\widehat{\tau_{HT}}(d_{10}, d_{00})\), which isolates
the effect of direct exposure in the absence of any interaction with
indirect exposure, as well as \(\widehat{\tau_{HT}}(d_{01}, d_{00})\),
which isolates the effect of indirect exposure in the absence of any
interaction with direct exposure. In this case, we cannot compute
\(\widehat{\tau_{HT}}(d_{11}, d_{00})\)---the interactive effect of
direct and indirect exposure---because in this small scale example no
unit received exposure \(d_{11}\).

\begin{Shaded}
\begin{Highlighting}[]
\NormalTok{yT_HT <-}\StringTok{ }\KeywordTok{estimates}\NormalTok{(D, observed_outcomes, prob_exposure, }\DataTypeTok{hop =} \DecValTok{1}\NormalTok{)}\OperatorTok{$}\NormalTok{yT_ht}
\end{Highlighting}
\end{Shaded}

The object \texttt{yT\_HT} is a named numeric vector which contains the
values of \(\widehat{y^T_{HT}}(d_k)\) for
\(d_k=d_{11}, d_{10}, d_{01}, d_{00}\) in that order. Therefore, in
order to compute \(\widehat{\tau_{HT}}\), we can take the difference of
each of these values with the value of exposure condition
\(d_{00}(\text{No Exposure})\) and then divide by the number of units
\(N\).

\begin{Shaded}
\begin{Highlighting}[]
\NormalTok{tau_HT <-}\StringTok{ }\NormalTok{((}\DecValTok{1}\OperatorTok{/}\NormalTok{N)}\OperatorTok{*}\NormalTok{(yT_HT}\OperatorTok{-}\NormalTok{yT_HT[}\StringTok{'no'}\NormalTok{])[}\KeywordTok{names}\NormalTok{(yT_HT)}\OperatorTok{!=}\StringTok{'no'}\NormalTok{])}
\NormalTok{tau_HT}
\end{Highlighting}
\end{Shaded}

\begin{verbatim}
dir_ind1 isol_dir     ind1 
      NA 41.61043 39.51643 
\end{verbatim}

In fact, the \texttt{estimates} function already computes
\(\widehat{\tau_{HT}}\) directly:

\begin{Shaded}
\begin{Highlighting}[]
\KeywordTok{estimates}\NormalTok{(D, observed_outcomes, prob_exposure, }\DataTypeTok{hop=}\DecValTok{1}\NormalTok{)}\OperatorTok{$}\NormalTok{tau_ht}
\end{Highlighting}
\end{Shaded}

\begin{verbatim}
dir_ind1 isol_dir     ind1 
      NA 41.61043 39.51643 
\end{verbatim}

The estimator \(\widehat{\tau_{HT}}(d_{k}, d_{l})\) is unbiased when is
estimated with \({\pi_i(d_k)}\) rather than \({\hat{\pi}_i(d_k)}\). As
mentioned above, when estimated with the latter, the estimator is not
unbiased, but the bias becomes negligible with a sufficiently large
number of replicates \(R\). We implement variance estimators for
\(\Var\left[\widehat{y^T_{HT}}(d_k)\right]\) and
\(\Var\left[\widehat{\tau_{HT}}(d_{k}, d_{l})\right]\) as derived in
Equation 11 of \citet{aronow2017estimating}. These are conservative
approximations to the exact variances that are guaranteed to have
non-negative bias relative to the variance of the randomization
distribution of the estimators. Because unbiased estimators for
\(\Var\left[\widehat{y^T_{HT}}(d_k)\right]\) are only identified when
the joint exposure probabilities of every pair of units is positive, it
is necessary to add a correction term to
\(\widehat{\Var}\left[\widehat{y^T_{HT}}(d_k)\right]\). Second, because
the
\(\Cov\left[\widehat{y^T_{HT}}(d_k), \widehat{y^T_{HT}}(d_l)\right]\) is
always unidentified, we use an approximation. Both of these corrections,
contribute to the non-negative bias of the variance estimator. We also
implement the constant effects variance estimator derived in
\citet{aronow2013dissertation}. This estimator operates under the
assumption that exposure effects do not vary across subjects, and
therefore \(y_i(d_k)=y_i(d_l) + \tau(d_{k}, d_{l})\) for every unit
\(i\). Then, one can estimate the variance by either plugging the
estimated \(\tau(d_{k}, d_{l})\) values into the expression of the
variance or using them to reconstruct the full schedule of potential
outcomes and then simulating new treatment assignments to approximate
the distribution of effect estimates. In the following applications, we
take the maximum between the constant effects variance estimator and the
conservative variance estimator developed in
\citet{aronow2017estimating}. Confidence intervals are based on a
large-N normal approximation:
\(\widehat{\tau_{HT}}(d_{k}, d_{l}) \pm z_{(1-\alpha)/2} \sqrt{\widehat{\Var}\left[\widehat{\tau_{HT}}(d_{k}, d_{l})\right]}\).

Asymptotic convergence of these estimators, and therefore the
reliability of normal approximations for inference, depend on whether
outcome dependence across units is limited. In particular, consistency
of \(\widehat{\tau_{HT}}(d_{k}, d_{l})\) follows from limits on the
amount of pairwise dependency in exposure conditions induced by both the
design and the exposure mapping as the sample size increases. Going back
to the antibullying program experiment of \citet{paluck2016changing},
this condition implies that as new students are added to a school, the
new peer connections that result between them an existing students
cannot be too extensive.

An alternative to the Horvitz-Thompson estimator is the Hajek estimator,
which improves efficiency with a small cost in terms of finite sample
bias. This estimator is a ratio approximation of the Horvitz-Thompson:
\[ \widehat{\mu}_H(d_k) = \frac{\sum_{i = 1}^{N} \I(D_i = d_k) \frac{Y_i}{\pi_i(d_k)}}{\sum_{i = 1}^{N} \I(D_i = d_k) \frac{1}{\pi_i(d_k)}}.\]

In the Horvitz-Thompson estimator \(\widehat{\mu_{HT}}(d_k)\) is high
variance because some randomizations yield units with extremely high
values of the weights \(1/\pi_i(d_k)\). The Hajek refinement allows the
denominator of the estimator to vary according to the sum of the weights
\(1/\pi_i(d_k)\), therefore shrinking the magnitude of the estimator
when its value is large, and increasing the magnitude of the estimator
when its value is small.

We extend the example developed above to consider a more realistic
sample size. In what follows here and in the following sections, we use
a set of \(N=400\) units and we randomly assign without replacement a
proportion \(p=0.1\) to treatment condition \(z_i=1\) with uniform
probability. As in the previous example, the network is modeled as
small-world with each unit directly connected on average to four other
units, and the potential outcomes follow the ``dilated effects''
scenario. We set the number of replications to \(R=10000\) to compute
exposure probabilities, and run 3000 simulated replications of the whole
experiment.

\begin{table}

\caption{\label{tab:tbhth}Comparing Horvitz-Thompson to Hajek Estimator, Using Approximate Exposure Probabilities}
\centering
\begin{threeparttable}
\begin{tabular}[t]{c|c|c|c|c|c|c|c}
\hline
Estimator & Estimand & True-Value & Average-Value & Bias & SD & RMSE & MeanSE\\
\hline
Horvitz-Thompson & $\tau(d_{11}, d_{00})$ & 58.12 & 57.33 & -0.78 & 45.32 & 45.32 & 45.94\\
\hline
Horvitz-Thompson & $\tau(d_{10}, d_{00})$ & 29.06 & 29.07 & 0.01 & 26.60 & 26.60 & 33.11\\
\hline
Horvitz-Thompson & $\tau(d_{01}, d_{00})$ & 14.53 & 14.66 & 0.13 & 10.45 & 10.45 & 10.33\\
\hline
Hajek & $\tau(d_{11}, d_{00})$ & 58.12 & 59.41 & 1.29 & 36.66 & 36.68 & 34.07\\
\hline
Hajek & $\tau(d_{10}, d_{00})$ & 29.06 & 28.92 & -0.14 & 21.55 & 21.55 & 25.09\\
\hline
Hajek & $\tau(d_{01}, d_{00})$ & 14.53 & 14.85 & 0.32 & 8.38 & 8.39 & 8.41\\
\hline
\end{tabular}
\begin{tablenotes}
\item \textit{Note: } 
\item Horvitz–Thompson estimator with maximum between conservative variance estimator and constant effects variance estimator. Hajek estimator with linearized variance estimator. True-Value = Value of estimand. Average-Value = Value of estimator. SD = Empirical standard deviation from simulation. RMSE = Root-mean-square error. MeanSE = mean standard error estimate. Estimators use approximate exposure probabilities calculated by drawing 10000 treatment assignments without replacement.
\end{tablenotes}
\end{threeparttable}
\end{table}

The result of the simulation shown in Table \ref{tab:tbhth} illustrates
that the Hajek estimator is more efficient than the Horvitz-Thompson
estimator. The empirical standard deviation from simulation is smaller
for the Hajek estimator, and the decrease in variance is with little
cost in bias as indicated by a smaller root-mean-square error. Moreover,
the variance estimator is consistent, given that the mean of the
standard error estimate approaches the empirical standard deviation from
simulation.

\subsection{Misspecified exposure
mappings}\label{misspecified-exposure-mappings}

We now consider the implications of misspecified exposure mappings.
Proposition 8.1 in \citet{aronow2017estimating} show what happens when
an exposure condition \(D_i = d_k\) specified by the experimenter is
actually consistent with multiple potential outcomes for unit \(i\), in
which case the exposure mapping is too coarse. Here, we examine these
issues in the context of our running example. We consider the case where
one assumes that interference happens only through direct peer
connections (first-degree interference), but in fact units are exposed
to treatment of their direct peers and to treatment of peers of their
direct peers (it is second-degree). We also consider what happens when
the experimenter ignores interference, but in fact it is first- or
second-degree.

Let us begin with an exposure mapping as above with four exposures for
first-degree interference (\(d_{11}\), \(d_{10}\), \(d_{01}\),
\(d_{00}\)), and then another exposure mapping with eight exposures
based on second-order interference (\(d_{111}\), \(d_{110}\),
\(d_{101}\), \(d_{100}\), \(d_{011}\), \(d_{010}\) \(d_{001}\),
\(d_{000}\)), and finally a no-interference exposure mapping with only
two exposures (\(d_{1}\), \(d_{0}\)). Then, suppose two possible true
data generating processes, with one exhibiting only first-degree
interference and the other exhibiting second-degree interference. The
exposure mapping is misspecified when the type of interference assumed
by the experimenter does not match the true data generating process.
This gives rise to six scenarios, two of which have correct exposure
mapping specifications and the rest being cases of misspecification.

To define a coherent notion of bias under misspecification, one needs to
define quantities of interest in terms of treatment regimes. In our
case, we consider the contrast between average outcomes under 100\%
treatment saturation versus 0\% saturation: \[
\tau(\mathbf{1},\mathbf{0}) = \frac{1}{N}\sum_{i=1}^Ny_i(\mathbf{1}) - y_i(\mathbf{0}),
\] where the \(\mathbf{1}\) and \(\mathbf{0}\) are meant to denote that
100\% or 0\% of units are assigned to treatment, respectively.

When the true data generating process involves no interference, then
\(\tau(\mathbf{1},\mathbf{0})\) is equivalent to the usual average
treatment effect (ATE). Under interference, this is not the case. When
the true data generating process involves only the first-degree
spillover as per our running example, then
\(\tau(\mathbf{1},\mathbf{0}) = \tau(d_{11}, d_{00})\). With
second-degree spillover,
\(\tau(\mathbf{1},\mathbf{0}) = \tau(d_{111}, d_{000})\).
Misspecification will result in working with inappropriate contrasts to
estimate \(\tau(\mathbf{1},\mathbf{0})\). For example, suppose the true
data generating process is first-degree, but one assumes no
interference. Then, one would mistakenly take the potential outcomes
\(y_i(d_{11})\) to be equivalent to \(y_i(d_{10})\), and use a mixture
of such outcomes in estimating the desired average of
\(y_i(\mathbf{1})\) outcomes. To see this, consider estimating the
population mean when everyone is treated, \(\mu(\mathbf{1})\), using a
Horvitz-Thompson estimator (which we use here for simplicity, results
for the Hajek estimator would be along the same lines). Then, if we
assume no interference, we would compute

\[
\begin{aligned}
\widehat{\mu_{HT, None}}(\mathbf{1}) & = \frac{1}{N}\sum_{i=1}^N \I(Z_i = 1)\frac{Y_i}{\pi_i} \\
& = \frac{1}{N}\sum_{i=1}^N \left(\I(D_i = d_{11})\frac{y_i(d_{11})}{\pi_i} + \I(D_i = d_{10})\frac{y_i(d_{10})}{\pi_i}\right),
\end{aligned}
\] where \(\pi_i = \text{Pr}[Z_i=1]\), while the unbiased estimator
would be, \[\begin{aligned}
\widehat{\mu_{HT}}(\mathbf{1}) = \widehat{\mu_{HT}}(d_{11}) & = \frac{1}{N}\sum_{i=1}^N \I(D_i = d_{11})\frac{Y_i}{\pi_i(d_{11})} \\
& = \frac{1}{N}\sum_{i=1}^N \I(D_i = d_{11})\frac{y_i(d_{11})}{\pi_i(d_{11})},
\end{aligned}\] where
\(\pi_i(d_{11}) = \text{Pr}[Z_i\I(\Z'\theta_i>0) = 1]\).

Therefore, the misspecified \(\widehat{\mu_{HT, None}}(\mathbf{1})\) is
biased for \(\widehat{\mu_{HT}}(\mathbf{1})\) insofar as
\(y_i(d_{11}) \ne y_i(d_{10})\) for some \(i\).

When the assumed exposure mapping considers higher-degree interference
than the true data-generating process, the resulting estimator can be
unbiased, but with a cost in variance. The reason is that this
misspecified estimator incorporates only a fraction of the available
units to construct the potential outcome average.

\begin{figure}

{\centering \includegraphics[width=.8\linewidth]{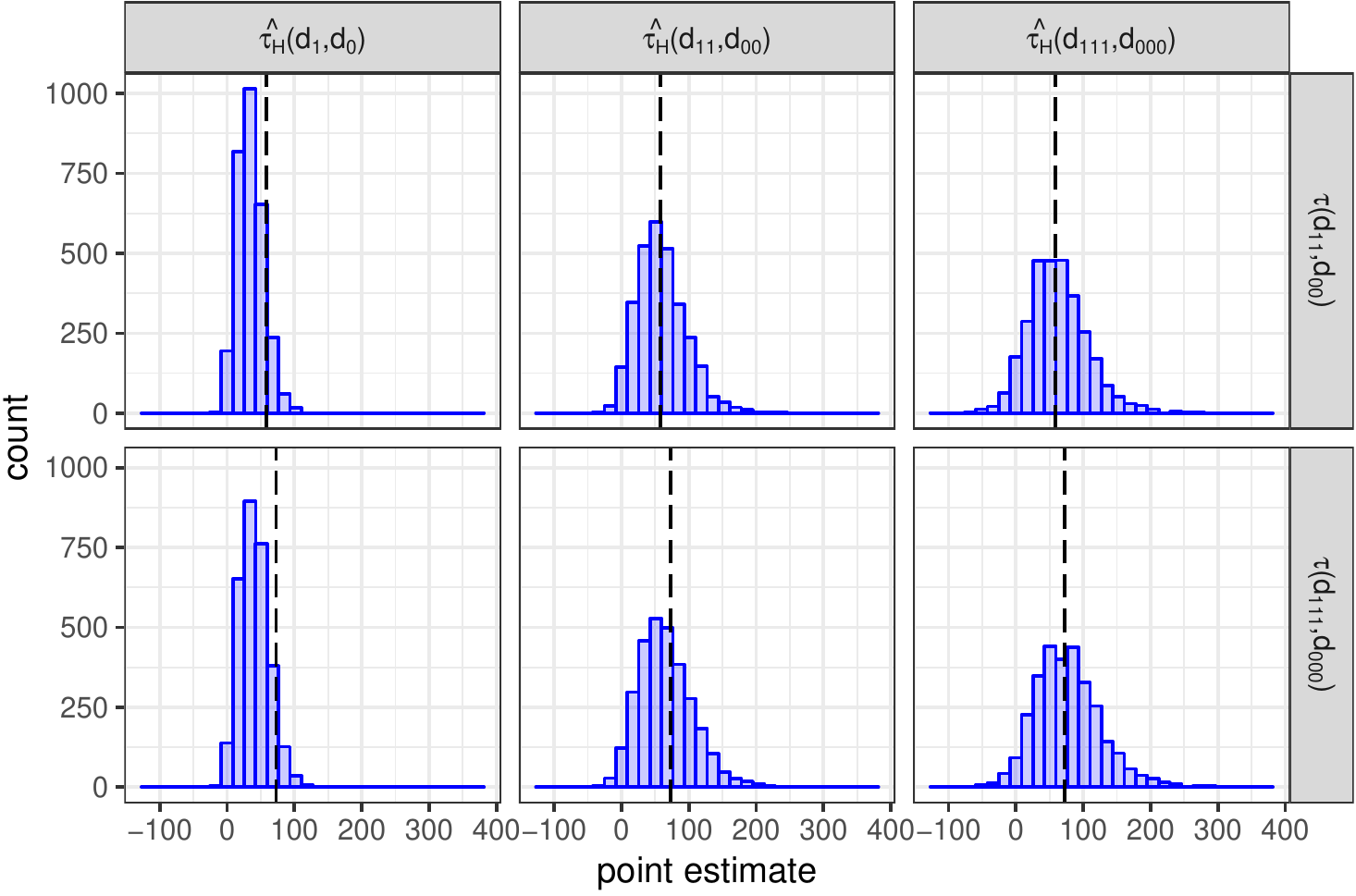} \includegraphics[width=.8\linewidth]{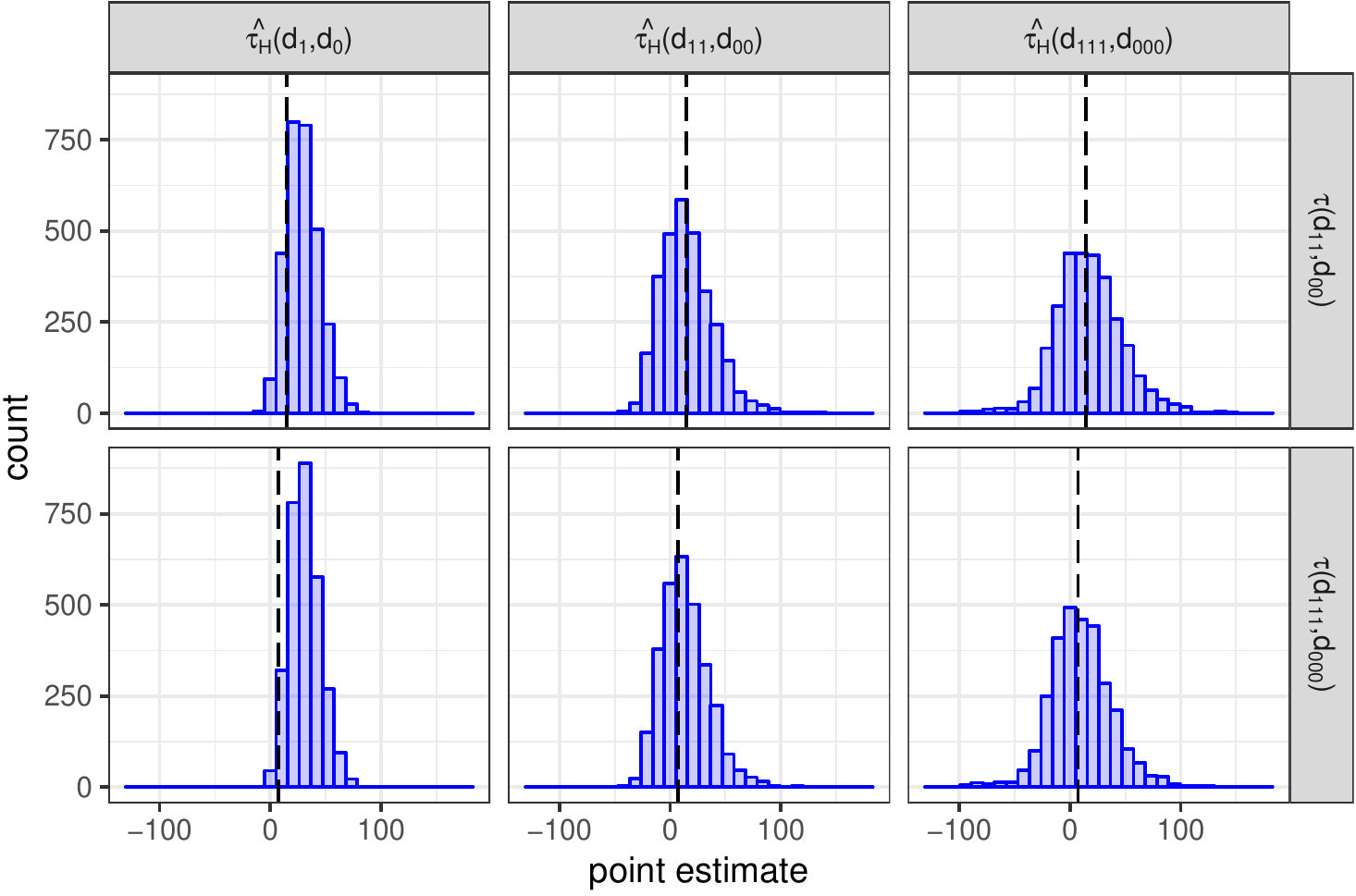} 

}

\caption{\label{fig:miss_exposure}From left to right the plot shows the distribution of the point estimates for 3000 simulations when the exposure mapping ignores interference, assumes first-degree, and second-degree interference, respectively. The top panel is for a case of positive spillover effects, and the bottom panel is for a case of negative spillover effects. In each panel, the upper-row DGP is first-degree interference and the lower-row DGP is second-degree interference. The dashed vertical line represents the true value of the target quantity, which is the average effect of going from no one treated to everyone treated.}\label{fig:unnamed-chunk-12}
\end{figure}

Table \ref{tab:tbposneg} and Figure \ref{fig:miss_exposure} illustrate
how estimates vary over these different forms of misspecification in our
simulated data. For the sake of completeness, we also present results
for another simulation where the spillover effects are negative
(essentially, choosing dilated effects multipliers when simulating
outcome data such that the multipliers for \(d_{11}\) and \(d_{01}\) are
smaller than those of \(d_{10}\) and \(d_{00}\), respectively). Looking
at Figure \ref{fig:miss_exposure}, from left to right we plot the
distribution of point estimates for \(\widehat{\tau}(d_{1}, d_{0})\)
(assuming no interference), \(\widehat{\tau_{H}}(d_{11}, d_{00})\)
(assuming first-degree interference), and
\(\widehat{\tau_{H}}(d_{111}, d_{000})\) (assuming second-degree
interference). Then, we vary whether the true data generating process
exhibits first- or second-degree interference. In all cases, the target
of inference is \(\tau(\mathbf{1}, \mathbf{0})\). Under first-degree
interference, \(\tau(\mathbf{1}, \mathbf{0}) =\tau(d_{11}, d_{00})\),
and under second-degree interference
\(\tau(\mathbf{1}, \mathbf{0})=\tau(d_{111}, d_{000})\). These
variations in data-generating processes are shown going up and down
Figure \ref{fig:miss_exposure}, for the positive spillovers case in the
top panel and the negative spillovers case in the bottom panel. True
values of the target quantities are shown with the dashed lines, and the
distributions of estimators over 3000 simulation runs are shown with the
histograms. We see that estimators are centered around the true
quantities when the exposure mapping incorporates equal- or
higher-degree interference than the true data generating process.
Because the estimator incorporates only a fraction of the available
units to construct the potential outcome average, the variance is higher
when the exposure mapping considers higher-degree interference than the
true data generating process. To the contrary, the estimators are biased
when the exposure mapping considers a lower-degree interference than the
true data generating process. In this case, as we explained above,
potential outcomes under different exposure conditions are taken to be
equivalent, and therefore, averaged together by the estimator when
constructing the potential outcome average. With positive spillover the
estimators' bias is negative (top panel), whereas the bias is positive
with a case of negative spillover (bottom panel). For monotonic
interference of the kinds considered here, bias is reduced more by
considering exposure mappings with more refined patterns of interference
(for example, no interference vs.~first-degree interference), as shown
in Theorem 2.3 of \citet{eckles2017design}. Table \ref{tab:tbposneg}
presents the evaluation metrics for each of the six scenarios depicted
in Figure \ref{fig:miss_exposure}: bias, standard deviation of simulated
estimates and root-mean-square error.

Note that this example uses a relatively small set of units (\(N=400\)),
and the unbiased estimators (in this case,
\(\widehat{\tau_{H}}(d_{11}, d_{00})\),
\(\widehat{\tau_{H}}(d_{111}, d_{000})\)) work with smaller subsets of
the data than the coarser, biased estimators (here,
\(\widehat{\tau_{H}}(d_{1}, d_{0})\)). This is apparent if one looks at
the standard deviations (SD) in Table \ref{tab:tbposneg}. As such, with
these sample sizes, the root-mean-square error is swamped by estimation
variance relative to bias. As \(N\) gets larger, this would change: the
SD would get smaller, but the bias would remain.

\begin{table}

\caption{\label{tab:tbposneg}Mispecifying Exposure Conditions}
\centering
\begin{threeparttable}
\begin{tabular}[t]{c|c|c|c|c|c|c}
\hline
Spillover & Estimand & Estimator & True-Value & Bias & SD & RMSE\\
\hline
positive & $\tau(d_{11}, d_{00})$ & $\hat{\tau_H}(d_{1}, d_{0})$ & 58.12 & -23.33 & 19.06 & 30.12\\
\hline
positive & $\tau(d_{11}, d_{00})$ & $\hat{\tau_H}(d_{11}, d_{00})$ & 58.12 & 1.29 & 36.66 & 36.68\\
\hline
positive & $\tau(d_{11}, d_{00})$ & $\hat{\tau_H}(d_{111}, d_{000})$ & 58.12 & 3.90 & 44.14 & 44.30\\
\hline
positive & $\tau(d_{111}, d_{000})$ & $\hat{\tau_H}(d_{1}, d_{0})$ & 72.65 & -32.21 & 21.42 & 38.68\\
\hline
positive & $\tau(d_{111}, d_{000})$ & $\hat{\tau_H}(d_{11}, d_{00})$ & 72.65 & -6.55 & 40.92 & 41.44\\
\hline
positive & $\tau(d_{111}, d_{000})$ & $\hat{\tau_H}(d_{111}, d_{000})$ & 72.65 & 4.28 & 48.95 & 49.13\\
\hline
negative & $\tau(d_{11}, d_{00})$ & $\hat{\tau_H}(d_{1}, d_{0})$ & 14.53 & 14.87 & 14.73 & 20.93\\
\hline
negative & $\tau(d_{11}, d_{00})$ & $\hat{\tau_H}(d_{11}, d_{00})$ & 14.53 & 0.85 & 23.36 & 23.37\\
\hline
negative & $\tau(d_{11}, d_{00})$ & $\hat{\tau_H}(d_{111}, d_{000})$ & 14.53 & 2.77 & 30.23 & 30.35\\
\hline
negative & $\tau(d_{111}, d_{000})$ & $\hat{\tau_H}(d_{1}, d_{0})$ & 7.26 & 23.92 & 13.65 & 27.54\\
\hline
negative & $\tau(d_{111}, d_{000})$ & $\hat{\tau_H}(d_{11}, d_{00})$ & 7.26 & 6.42 & 21.27 & 22.21\\
\hline
negative & $\tau(d_{111}, d_{000})$ & $\hat{\tau_H}(d_{111}, d_{000})$ & 7.26 & 2.58 & 28.05 & 28.16\\
\hline
\end{tabular}
\begin{tablenotes}
\item \textit{Note: } 
\item Hajek estimator with linearized variance estimator. True-Value = Value of estimand. SD = Empirical standard deviation from simulation. RMSE = Root-mean-square error. Estimators use approximate exposure probabilities computed by drawing 10000 treatment assignments without replacement from the set of possible treatment profiles.
\end{tablenotes}
\end{threeparttable}
\end{table}

\subsection{Misspecified network ties}\label{misspecified-network-ties}

Another type of misspecification is when the network measured by the
experimenter is different than the actual interference network. In this
case, the bias of the estimator will increase with the proportion of
mis-measured ties. For example, with positive spillover and an elicited
network with missing ties, one would mistakenly take a mixture of the
potential outcomes \(y_i(d_{01})\) and \(y_i(d_{00})\) in estimating the
average of \(y_i(d_{00})\), leading to an overestimate and thereby
contributing negative bias for any estimate of an effect relative to
\(\mu(d_{00})\).

\begin{figure}

{\centering \includegraphics{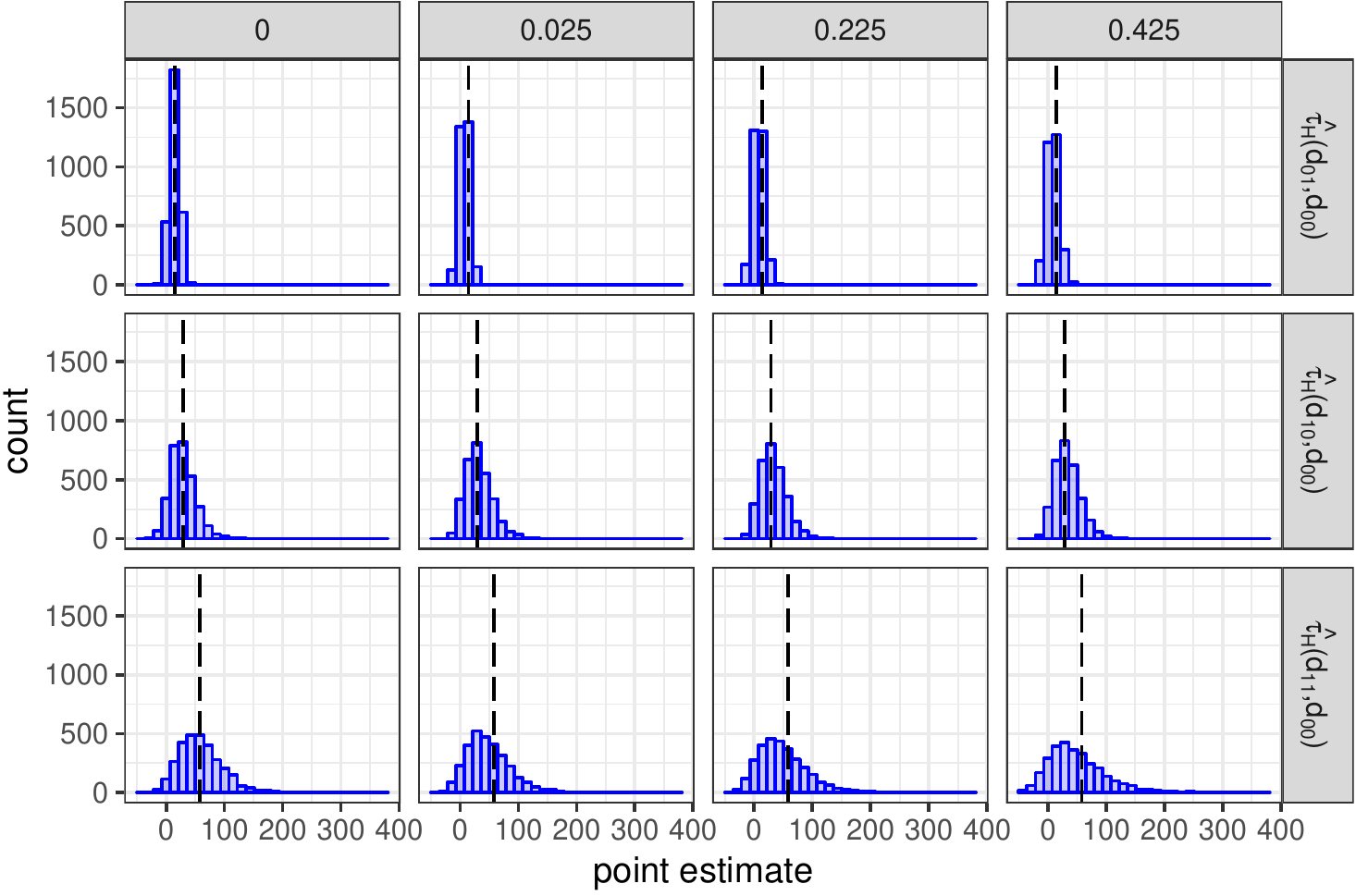} 

}

\caption{\label{fig:miss_ties}The plot shows the distribution of point estimates for 3000 simulations given different proportions of missing ties for a case of positive spillover. The dashed vertical line represents the true value of the target quantity.}\label{fig:unnamed-chunk-13}
\end{figure}

Figure \ref{fig:miss_ties} shows more general consequences of working
with network estimates based on a network that is randomly missing
varying proportions of ties in the interference network. The data
generating process is the first-degree interference set-up described
above and the figure shows results from 3000 simulation draws, plotting
distributions of point estimates for
\(\widehat{\tau_{H}}(d_{01}, d_{00})\),
\(\widehat{\tau_{H}}(d_{10}, d_{00})\), and
\(\widehat{\tau_{H}}(d_{11}, d_{00})\). The distribution of point
estimates is centered around the target quantity (the dashed line) when
the proportion of unmeasured ties is zero, indicating unbiasedness.
However, as the proportion of unmeasured ties increases the
distribution's shift to the left or right, depending on the net effect
of the biases for the estimated potential outcome means.

\subsection{Sensitivity analysis for
misspecification}\label{sensitivity-analysis-for-misspecification}

\citet{egami2017unbiased} proposes a sensitivity analysis for
misspecification of the exposure mapping due to unobserved networks. The
sensitivity analysis considers a situation where spillover effects occur
on an unobserved ``offline'' network when the experimenter only observes
an ``online'' network. The case captures situations where one uses, for
example, online social network data like a Twitter follower network to
specify the interference network, but that interference can also occur
via offline ties like a network of high school classmates that is not
captured by the online network. The analysis could apply to any
situation where the measured network fails to capture all relevant ties
in the true interference network. The sensitivity analysis focuses on
estimating the \emph{average network-specific spillover effect} (ANSE),
which is the average causal effect of changing the treatment status of
neighbors in the online network (for example, status of Twitter ties),
without changing the treatment status of neighbors in the offline
network (high school classmates). The analysis also requires the
\emph{stratified interference} assumption (to which we return below),
which assumes that potential outcomes of unit \emph{i} are affected by
\emph{i}'s own treatment assignment and only the treated
\emph{proportion} of online and offline neighbors; the precise set of
treated neighbors does not matter. Under this assumption,
\citet{egami2017unbiased} develops parametric and nonparametric
sensitivity analysis methods. The parametric method in addition assumes
that the unobserved network spillover effect is linear and additive, and
helps to derive simple formal conditions under which unobserved networks
would explain away the estimated ANSE. The nonparametric method assumes
instead that outcomes are just non-negative, and is used to bound the
ANSE.

\subsection{Efficient designs to estimate effects under
interference}\label{efficient-designs-to-estimate-effects-under-interference}

We now consider implications of the preceding analysis for designing
experiments that estimate exposure-specific effects efficiently. Let us
first consider why such designs are needed. If the goal is to estimate
the average difference in unit potential outcomes under 100\% versus 0\%
treatment saturation, defined as \(\tau(\mathbf{1},\mathbf{0})\) above,
naive designs can perform poorly. \citet{ugander2013graph} show that
unit-level designs need not even yield asymptotically consistent
estimators under first-degree interference if degree is also increasing
as the sample size does. The problem is that very few units end up with
either all or no first-degree neighbors treated.

Analogously to designs for the partial interference setting considered
below, \citet{ugander2013graph} propose cluster-randomized designs. In
this \emph{graph cluster randomization} the network (or graph) is
partitioned into a set of clusters, such that units closer to each other
in the network are assigned to the same cluster, and then treatment
randomization is performed at the cluster level. Estimation then employs
the inverse-probability weighted methods described above. When
\(\tau(\mathbf{1},\mathbf{0})\) is the estimand, graph cluster
randomization can lead to exponentially (in sample size) lower estimator
variance as compared to unit-level random assignment. This is because
under graph cluster randomization, connected units are assigned to the
same treatment condition more often than would happen with unit level
assignment, increasing the expected number of units who are exposed to
one of the full neighborhood exposure conditions. On the other hand,
assigning units by cluster can contribute to increases in variance,
especially insofar as it means that---due to homophily---units with
similar outcomes will tend to be assigned to one or another exposure
condition together. \citet{ugander2013graph} analyze an intuitive graph
clustering method, \(\epsilon\)-net clustering, where clusters are
formed by finding a set of units such that all units in the set are at
least \(\epsilon\) hops of each other, and every unit outside the set is
within \(\epsilon-1\) hops of a unit in the set. With \(\epsilon=3\)
graph cluster randomization has some desirable asymptotic properties
even as average degree is growing with the sample size. In practice,
experimenters can use many other methods for graph partitioning or
community detection, but these are more difficult to study analytically;
\citet{saveski2017detecting} includes empirical comparison of such
methods.

We compare the Horvitz-Thompson and Hajek estimators and their variance
for the causal estimand of full neighborhood exposure,
\(\tau(d_{\mathbf{1}},d_{\mathbf{0}})\), for the case in which units are
assigned to treatment via unit-level randomization as opposed to 3-net
clustering randomization. In both cases, our simulation assumes that
treatment assignment has a \(\text{Bernoulli}(1/2)\) distribution, there
are 400 units connected by a small world network with average degree 4,
and units respond to treatment following a dilated effects scenario in
which the outcomes have a positive lower bound and are correlated with
the unit's first and second order degree. To compute the exposure
probabilities, we set the number of replications to \(R=10000\). We run
3000 simulated replications of the experiment. The results of the
simulation are presented in Table \ref{tab:tbunitcluster}. They show
that graph cluster randomization leads to substantially lower estimator
variability with no cost in bias as indicated by a smaller
root-mean-square error.

On the other hand, if the exposure model is misspecified, neither design
yields unbiased or consistent estimators; however, under some models in
which interference is monotonic, graph cluster randomization can
nonetheless reduce bias at the cost of variance
\citep{eckles2017design}.

The above discussion considers contrasting 100\% versus 0\% treatment
saturation, for which graph cluster randomization---and other designs
that produce network autocorrelation in treatment---can be advantageous.
Other estimands may motivate quite different experimental designs. For
example, if the goal is to estimate interference effects (e.g.,
\(\tau(d_{01}, d_{00})\)), graph clustered randomization with a perfect
partitioning of a network into its \(k\) connected components may make
estimation impossible, since all units will have the same treatment as
their neighbors.

\begin{table}

\caption{\label{tab:tbunitcluster}Comparing Unit to Cluster Randomization}
\centering
\begin{threeparttable}
\begin{tabular}[t]{c|c|c|c|c|c|c|c}
\hline
Randomization & Estimator & Estimand & True-Value & Average-Value & Bias & SD & RMSE\\
\hline
Unit & Horvitz-Thompson & $\tau(d_{\mathbf{1}}, d_{\mathbf{0}})$ & 58.12 & 55.70 & -2.42 & 139.65 & 139.65\\
\hline
Cluster & Horvitz-Thompson & $\tau(d_{\mathbf{1}}, d_{\mathbf{0}})$ & 58.12 & 57.52 & -0.60 & 61.31 & 61.30\\
\hline
Unit & Hajek & $\tau(d_{\mathbf{1}}, d_{\mathbf{0}})$ & 58.12 & 48.27 & -9.85 & 52.89 & 53.79\\
\hline
Cluster & Hajek & $\tau(d_{\mathbf{1}}, d_{\mathbf{0}})$ & 58.12 & 55.01 & -3.10 & 28.89 & 29.05\\
\hline
\end{tabular}
\begin{tablenotes}
\item \textit{Note: } 
\item Unit = unit-level randomization. Cluster = 3-net clustering. Horvitz–Thompson estimator with maximum between conservative variance estimator and constant effects variance estimator. Hajek estimator with linearized variance estimator. True-Value = Value of estimand. Average-Value = Value of estimator. SD = Empirical standard deviation from simulation. RMSE = Root-mean-square error. Estimators use approximate exposure probabilities computed by drawing 10000 treatment assignments without replacement.
\end{tablenotes}
\end{threeparttable}
\end{table}

\subsection{Empirical Studies}\label{empirical-studies}

\citet{bond201261} analyze the influence of peers within a large scale
voter-mobilization network experiment delivering messages to 61 million
Facebook users during the day of the 2010 U.S. Congressional Election.
Users were randomly assigned to a social message (98\% of users), an
informational message (1\%) or a control condition (1\%). The two
treatment conditions encouraged users to vote, provided a polling-place
link, allow them to express they had voted by clicking an ``I voted''
button and showed a counter with the number of users who had previously
reported voting. In addition, the social message presented pictures of
friends who had reported voting. The polling-place link was used as a
measure of desire to seek information about the election and the ``I
voted'' button as self-reported voting. Additionally, the turnout of
about 10\% of the users was validated with public voting records. The
estimated results suggest positive direct effects of the social message
on information seeking, self-reported voting and validated voting,
whereas the informational message did not affect turnout. Results
suggest positive spillover effects on ``close friends'' with whom they
interact frequently on Facebook and who are likely to have real-world,
face-to-face relationships, but no effects on distant friends. The way
in which \citet{bond201261} analyzed spillover effects did not properly
account for direct effects, a problem discussed by
\citet{taylor-eckles-social-influence-experiments} when considering
sharp null hypothesis tests on indirect effects.

A follow up study was conducted by \citet{jones2017social} during the
2012 Presidential Election. In this study, the experimental conditions
were adjusted to better understand the mechanisms that were likely to
drive effects and the analysis of spillover effects corrected the
problems with \citet{bond201261}. Instead of treating the large majority
of users with the social message condition and assigning the rest to an
informational or control conditions, this time a 2x2 design varied
whether or not individuals saw a post at the top of their News Feeds
that encouraged turnout in a similar way than the social message (the
banner condition), and whether or not users saw individual posts within
their News Feed regarding friends' voting if at least one of their
friends in the banner condition had clicked on the ``I voted'' button
(the feed condition). In the 2010 experiment, users in the banner
condition also saw these messages within their feeds, not allowing to
separate the encouragement from the social effect. Findings suggest that
users directly exposed to both the banner and feed conditions were
significantly more likely to have voted than those in the control
condition. Spillover effects happened through those encouraged to vote
with the banner condition, whereas the feed treatment did not spill over
to close friends. However, this time there were other differences in the
messaging due to the context of the 2x2 design. For example, messages in
the feed condition did not contain the button to self-report voting or
the link to find the polling place. These features would also need to be
randomized to learn more about the differences between indirect effects
across the banner and feed conditions.

\citet{coppock2016treatments} explore direct and spillover effects of a
mobilization campaign on Twitter on informal political participation,
particularly signing an online petition and sharing the petition link.
Followers of an environmental nonprofit advocacy organization were
randomly assigned to receive a private (2/3 of followers) or a public
message (1/3), and those receiving the private message were either
primed to have an identity of high commitment (organizer) or of low
commitment (follower) in equal shares. A second manipulation encouraged
a random subset of petition signers to share the petition with their own
followers (who are also followers of the environmental organization and
who mostly follow one petition signer) as a way to measure spillover
effects among those who actually respond directly to treatment. The
design of this second manipulation reduces the number of exposure
conditions that would otherwise be prohibitively large when the
researcher is interested in measuring indirect effects for every
possible number of treated users who are followed (from 0 to 601,
according to the largest units' degree in this network), while
accounting for exposure probabilities, and without parameterizing the
response to exposure in the exposure mapping. Results show that direct
messages boost petition signatures and tweet behavior, and that priming
the follower identity is more effective than the organizer identity.
Regarding indirect effects there is evidence that signing the petition
was influenced by others' treatment.

Recent empirical studies exploiting offline networks include
\citet{green2016effects} who analyze spatial spillover effects in a
series of field experiments testing the impact of lawn signs on vote
outcomes by planting them in randomly selected voting precincts. In this
case, to account for indirect effects the experimental design ensured
that two neighboring precincts would not be assigned to direct treatment
at the same time. \citet{baicker2005spillover} exploits exogenous shocks
to state medical spending to explore if spending decisions spill over to
neighboring states, while \citet{isen2014local} leverages a
discontinuity from local referendum results to assess if fiscal
decisions of one jurisdiction, particularly taxing and spending,
influence the fiscal decisions of its neighbors, and
\citet{rogowski2012estimating} use the House office lottery (in which
newly elected members select their office spaces in a randomly chosen
order) as an instrumental variable to estimate the impact of legislative
networks on roll call behavior and cosponsorship decisions. The landmark
study by \citet{sacerdote2001peer} uses natural random assignment of
college roommates to measure spillover effects on educational
performance outcomes.

Other studies have focused on strategies for targeting interventions in
networks (i.e., seeding) so as to capitalize on heterogeneous
spillovers. While there is a large theoretical and algorithmic
literature on this problem (influence maximization), triggered by
\citet{kempe2003maximizing} who provide a set of algorithms to maximize
behavior diffusion when the researcher has knowledge of the network,
there are only a few randomized experiments. These typically rely on
imposing partial interference assumptions (see below) so that outcomes
of different, villages or schools, for example, can be treated as
independent observations. Some studies used random or haphazard
assignment of treatment to analyze what types of units produce the
largest spillovers. For example, the previously mentioned antibullying
program study by \citet{paluck2016changing}, measured the network
structure of 56 schools in New Jersey to analyze peer diffusion effects
of randomly selected seed groups of students encouraged to take a stance
against conflict at school, finding that students with more direct
connections are the most effective at influencing social norms and
behavior among their direct peers and at the school-level. Similarly,
\citet{banerjee2013diffusion} study the impact of the (non-randomized)
choice of targeted individuals in the diffusion of participation in a
new micro-finance loan program in India that invited leaders to an
informational meeting and asked them to spread information about the
loans. The authors develop a model of word-of-mouth diffusion and apply
it to network data of 43 villages, which was collected by surveying
households before the start of the program. The model distinguishes
between information passing (learning about the program from neighbors)
and endorsement (being influenced by neighbors' adoption of the
program---what we refer to in this chapter as contagion). This allows
one to tease out the likelihood of information passing through
participants as opposed to non-participants, and the marginal
endorsement effect conditional on being informed. These estimates are
used to propose measures of individuals' effectiveness as seeding
points. A smaller number of studies use experiments designed to compare
seeding strategies. For example, \citet{kim2015social} compare the
effectiveness of three seeding strategies: randomly selected
individuals, individuals with the highest number of direct connections,
and random friends from a nominated set of friends of random individuals
(one-hop targeting), on take-up rates of a public health program in
rural municipalities in Honduras; relying on strong parametric and
independence assumptions, they find that, for one of two behaviors,
one-hop targeting performs best. A similar strategy is examined by
\citet{banerjee2019using}, but with ``ambitious'' questions that ask
respondents to select someone who would be good at spreading
information; the results provide suggestive evidence that this strategy
may increase spread over random seeding and seeding using village
leaders. Other field experiments have been conducted for diffusion of
agricultural knowledge and technology
\citep{beaman2018diffusion, beaman2018can}. \citet{chin2018evaluating}
present estimators and optimal experimental designs for studying seeding
strategies that make use of at-most partial network information, as do
strategies studied by \citet{kim2015social} and
\citet{banerjee2019using}.

\section{Partial Interference and Marginal Causal
Effects}\label{partial-interference-and-marginal-causal-effects}

The analysis in the preceding section is quite general in that it does
little to restrict the network over which interference can occur. The
only restrictions was a ``local interference'' assumption needed for
asymptotic results to hold. The drawback, however, was that one needed
to have a specification of the interference networks that was either
complete or that overcompensated for interference in the true data
generating process. In this section, we review the approach of
\citet{hudgens2008toward}, who work under the assumption that the
interference network per se is unknown, however one can assume that
interference is limited to occurring within well-defined and
non-overlapping groups, for example, villages or households. The
assumption that interference does not cross group boundaries is known as
``partial interference'' \citep{sobel2006}. If the interference network
is not known, then we cannot map each assignment vector \(\mathbf{z}\)
to an exposure, in which case we cannot estimate exposure-specific
effects. Rather, \citet{hudgens2008toward} define more agnostic
``marginal causal effects'' that average over sets of assignment
profiles that could, in some unspecified way, generate spillover
effects. This will be made more precise below. The analysis depends on a
two-stage hierarchical design, where groups are first randomly assigned
to a level of treatment saturation, and then units within groups are
randomly assigned to treatment with probability equal to their group
saturation rate.

\subsection{Marginal Causal Effects}\label{marginal-causal-effects}

In the most general case, each treatment assignment \(\z \in \Omega\)
generates a distinct potential outcome for unit \(i\). Under partial
interference, the potential outcome for unit \(i\) depends only on the
treatment assignments for units in \(i\)'s group \(g\). For example,
suppose that we have six units, \(i=1,2,3,4,5,6\), split up into two
groups such that group A contains units 1, 2, and 3, and group B
contains units 4, 5, and 6. Assuming that unit 2 in group A was assigned
to treatment, partial interference would imply that unit 1's potential
outcomes would be the same for all assignment vectors in the set

\begin{align*}
\{(0,1,0,\mathbf{z}_{i,B}):\mathbf{z}_{i,B}\in \Omega_B \} = \{ & (0,1,0,0,0,0), (0,1,0,1,0,0), (0,1,0,0,1,0), (0,1,0,0,0,1), \\
& (0,1,0,1,1,0), (0,1,0,1,0,1), (0,1,0,0,1,1), (0,1,0,1,1,1)\},
\end{align*}

where \(\Omega_B\) is the set of possible assignments for group B. At
the same time, it may very well be that assigning unit 3 to treatment
instead of unit 2 would have different implications for unit 1's
potential outcomes---i.e., it may be that
\(y_{1,A}(0,1,0,\mathbf{z}_{i,B}) \ne y_{1,A}(0,0,1,\mathbf{z}_{i,B})\).
Moreover, it would be safe to assume that either of these might differ
from unit 1's outcome if no one in group A were assigned to treatment,
\(y_{1,A}(0,0,0,\mathbf{z}_{i,B})\), or if both 2 and 3 were assigned to
treatment \(y_{1,A}(0,1,1,\mathbf{z}_{i,B})\). Now, suppose three fair
coin flips were used to determine whether unit 1, unit 2, or unit 3
should be assigned to treatment. Then, there is a 50-50 chance that unit
1 would be assigned to control. Conditional on unit 1 being assigned to
control, the expected value of unit 1's outcome would be the average
over the four potential outcomes \(y_{1,A}\) enumerated above. This
expected value is unit 1's marginal (i.e., average) potential outcome
given that unit 1 is not treated but under a regime that assigns units
in 1's group to treatment with 50-50 probability.
\citet{hudgens2008toward}'s analysis defines marginal causal effects as
contrasts between such marginal potential outcomes.

More generally, we refer to the individual's marginal potential outcome
as \(y_{ig}(z; \psi)\) when \(i\) is assigned to treatment value \(z\)
and other treatment assignments are determined by an assignment regime
characterized by the parameter \(\psi\), which describes the degree of
treatment saturation. In the simple example in the preceding paragraph,
we have \(\psi = 0.50\) to describe the regime where each unit is
assigned to treatment using a Bernoulli draw with \(p=0.50\), in which
case the expected saturation is 50\%. Under complete random assignment,
which fixes the number of treated and control units, \(\psi\) could
index the share of units assigned to treatment.

\citet{hudgens2008toward} consider four types of marginal causal
effects: direct, indirect, total, and overall effects. The direct effect
for a particular unit corresponds to the difference between the unit's
potential outcomes when its treatment assignment changes while the group
treatment assignment is kept fixed to a given saturation. Then, the
\emph{group average direct causal effect}, under treatment saturation
\(\psi\) and group size \(n_g\) can be defined as
\(\tau^D_g(\psi) = \frac{1}{n_g} \sum_{i = 1}^{n_g} y_{ig}(1; \psi) - \frac{1}{n_g} \sum_{i = 1}^{n_g} y_{ig}(0; \psi)\),
and the \emph{population average direct causal effect} \(\tau^D(\psi)\)
(or simply, the average direct causal effect) is the average of
\(\tau^D_g(\psi)\) across groups. The indirect effect describes the
effect on a unit of the treatment received by others in the group, and
is obtained from differences across saturation values \(\psi\) and
\(\phi\). Thus, the \emph{group average indirect causal effect} is
\(\tau^I_g(\psi, \phi) = \frac{1}{n_g} \sum_{i = 1}^{n_g} y_{ig}(0; \psi) - \frac{1}{n_g} \sum_{i = 1}^{n_g} y_{ig}(0; \phi)\),
while the average indirect causal effect averages across groups. The
total causal effect combines the direct and indirect effects to capture
the effect of being directly treated and exposed to the treatment by
others in the group. Thus, the \emph{group average total causal effect}
is
\(\tau^{To}_g(\psi,\phi) = \frac{1}{n_g} \sum_{i = 1}^{n_g} y_{ig}(1; \psi) - \frac{1}{n_g} \sum_{i = 1}^{n_g} y_{ig}(0; \phi)\),
and the average total causal effect takes the average of
\(\tau^{To}_g(\psi,\phi)\) across groups. Under no-interference, the
indirect causal effect is zero and the total causal effect equals the
direct causal effect. Finally, the overall causal effect corresponds to
the group's response to different treatment saturation levels. The
\emph{group average overall causal effect} can be written as
\(\tau^{O}_g(\psi,\phi) = \frac{1}{n_g} \sum_{i = 1}^{n_g} y_{ig}(\psi) - \frac{1}{n_g} \sum_{i = 1}^{n_g} y_{ig}(\phi)\),
where \(y_{ig}(\psi)\) is the average potential outcome for unit \(i\)
in group \(g\) under all possible assignments given saturation \(\psi\).
We average the \(\tau^{O}_g(\psi,\phi)\) across groups to obtain the
average overall causal effect.

Under Bernoulli random assignment, these causal quantities are well
defined and have a clean \emph{ceteris paribus} interpretation. Under
any other design, the causal interpretation is not always clean. For
example, under complete random assignment the individual-level direct
effect for unit \(i\) would contrast outcomes with different numbers of
group members other than \(i\) assigned to treatment (i.e.,
\(\psi(n_g-1)\) when \(i\) is assigned to treatment and \(\psi(n_g)\)
when \(i\) is assigned to control). This point is discussed by
\citet{savje2017average}.

Let the observed outcome be \(Y_{ig}=y_{ig}(\Z_{ig})\) under the group
treatment assignment \(\Z_{ig}\) and suppose treatment saturation is
\(\psi\). Given Bernoulli assignment with probability \(\psi\), an
unbiased estimator for \(\frac{1}{n} \sum_{i = 1}^{n} y_{ig}(z; \psi)\),
with \(z=\{0,1\}\), is
\[\widehat{y}_{g}(z; \psi) = \frac{\sum_{i = 1}^{n} Y_{ig}\I(Z_{ig}=z)}{\sum_{i = 1}^{n}\I(Z_{ig}=z)},\]
and for the population average potential outcome, it is
\[\widehat{y}(z; \psi) = \frac{\sum_{g = 1}^{G} \widehat{y}_{g}(z; \psi)\I(S_g = \psi)}{\sum_{g = 1}^{G}\I(S_g = \psi)},\]
where \(S_g\) is the saturation level of group \(g\) (conditionally on
the denominators of these estimators being nonzero). Therefore, an
unbiased estimator for the average direct effect is
\(\widehat{\tau}^D(\psi)=\widehat{y}(1; \psi) - \widehat{y}(0; \psi)\).
Unbiased estimators for the other causal estimands of interest are
defined analogously:
\(\widehat{\tau}^I(\psi, \phi) = \widehat{y}(0; \psi) - \widehat{y}(0; \phi)\),
\(\widehat{\tau}^{To}(\psi, \phi) = \widehat{y}(1; \psi) - \widehat{y}(0; \phi)\),
and
\(\widehat{\tau}^O(\psi, \phi)=\widehat{y}(\psi) - \widehat{y}(\phi)\),
where \(\widehat{y}(\psi)\) is the average across groups assigned to
saturation \(\psi\) of the average observed outcomes of units in the
group under treatment assignment \(\Z_{ig}\).

For inference, \citet{hudgens2008toward} derive variance estimators
under an assumption called stratified interference. This refers to the
situation where potential outcomes for unit \(i\) in group \(g\) do not
vary on the basis of which other units in group \(g\) are assigned to
treatment, only in the number or share of other units assigned to
treatment. This assumption reduces the problem statistically to the
usual stratified setting without interference. When stratified
interference holds, their proposed variance estimators are unbiased if
unit causal effects are additive (e.g.,
\(y_{ig}(1,\psi) = y_{ig}(0, \psi) + d\), where \(d\) is constant), and
otherwise positively biased. \citet{liu_hudgens2014} discuss conditions
for asymptotic normality. \citet{tchetgen2012causal} extend
\citet{hudgens2008toward}'s results, providing conservative variance
estimators, a framework for finite sample inference with binary
outcomes, and extensions to observational studies. \citet{liu2014large}
develop asymptotic results for two-stage designs.

Both \citet{sinclair2012detecting} and \citet{baird2017optimal} discuss
statistical power for hierarchical designs. \citet{baird2017optimal}
offer thorough consideration of the optimal choice (in terms of
statistical power) of saturation levels (\(\psi\) and \(\phi\)) and a
distribution of these levels over groups for estimating direct,
indirect, total, and overall effects. Their methods assume the
population is partitioned into equal-sized non overlapping groups, that
partial interference and stratified interference hold, and a
linear-in-means outcome model. The optimal set of saturations and shares
of groups assigned to each saturation depends on the correlation of
potential outcomes within groups.

We show with a toy example how to compute these estimators and their
variance. Suppose we have 18 units equally divided in 6 groups. In the
first stage half of the groups are assigned to saturation \(\psi\) and
the other half to \(\phi\) with equal probability. In the second stage,
using complete random assignment two-thirds of units in groups with
saturation \(\psi\) are assigned to the treatment condition and
one-third to the control condition, and one-third of units in groups
with saturation \(\phi\) are assigned to treatment, while two-thirds to
control. For the purposes of the simulation, we will assume stratified
interference and compute potential outcomes under a dilated effects
scenario such that \(y_{ig}(1, \psi)=2 \times y_{ig}(0, \phi)\),
\(y_{ig}(1, \phi)=1.5 \times y_{ig}(0, \phi)\), and
\(y_{ig}(0, \psi)=1.25 \times y_{ig}(0, \phi)\), where
\(y_{ig}(0, \phi)\) is obtained with a random draw from an absolute
standard normal distribution.

We display the structure of the post-treatment data that the
experimenter has to have to compute the estimators. (We show the data of
six units only.) In this case, the realized saturation for \emph{group
1} is \(\psi\) (the group is assigned to the treatment condition in the
first stage as indicated in column ``group\_tr'' with value 1, and 2/3
of units within the group are treated in the second stage as shown in
column ``indiv\_tr'') and for \emph{group 4} is \(\phi\) (the group is
assigned to the control condition in the first stage, and therefore 1/3
of units are treated in the second stage):

\begin{Shaded}
\begin{Highlighting}[]
\NormalTok{post_tr_data[}\KeywordTok{c}\NormalTok{(}\DecValTok{1}\OperatorTok{:}\DecValTok{3}\NormalTok{,}\DecValTok{10}\OperatorTok{:}\DecValTok{12}\NormalTok{),]}
\end{Highlighting}
\end{Shaded}

\begin{verbatim}
   group group_tr indiv_tr obs_outcome
1      1        1        0   1.9269359
2      1        1        1   0.2788864
3      1        1        1   0.9606388
10     4        0        1   0.9062178
11     4        0        0   1.0599419
12     4        0        0   0.6051009
\end{verbatim}

The estimators and their variance are computed under the stratified
interference assumption with the following function from the
\texttt{interference} package:

\begin{Shaded}
\begin{Highlighting}[]
\NormalTok{estimates <-}\StringTok{ }\KeywordTok{estimates_hierarchical}\NormalTok{(post_tr_data)}
\NormalTok{causal_effects <-}\StringTok{ }\KeywordTok{unlist}\NormalTok{(estimates[[}\DecValTok{1}\NormalTok{]])}
\NormalTok{variance <-}\StringTok{ }\KeywordTok{unlist}\NormalTok{(estimates[[}\DecValTok{2}\NormalTok{]])}
\NormalTok{causal_effects}
\end{Highlighting}
\end{Shaded}

\begin{verbatim}
direct_psi_hat direct_phi_hat   indirect_hat      total_hat    overall_hat 
     1.0009851      0.2318532      0.2195894      1.2205745      0.8096284 
\end{verbatim}

\begin{Shaded}
\begin{Highlighting}[]
\NormalTok{variance}
\end{Highlighting}
\end{Shaded}

\begin{verbatim}
var_direct_psi_hat var_direct_phi_hat   var_indirect_hat 
        0.71255260         0.04854851         0.28000088 
     var_total_hat    var_overall_hat 
        0.88936810         0.39005445 
\end{verbatim}

\subsection{Misspecifying partial
interference}\label{misspecifying-partial-interference}

What happens when the experimenter assumes partial interference but in
fact there is interference not only within groups, but also across
groups? We show that when partial interference does not hold, the
proposed estimators for the direct, indirect, total and overall causal
effects are biased.

We assume a scenario in which groups belong to tracts and interference
happens within groups and across groups within tracts, but not across
tracts. Each tract is composed of two groups. The sample size is 450
units, equally divided into six groups. As before, units are assigned to
treatment under a two-stage randomized experiment, with treatment
saturation set to \(\psi=2/3\) and baseline saturation to \(\phi=1/3\).
We assume stratified interference at the level of \emph{tracts} to
compute unit potential outcomes. We continue to use a dilated effects
scenario, but in this case it is the proportion of treated units in the
tract that matters, as opposed to the proportion of treated units in the
group.

Causal estimands are in this case defined as differences between unit
potential outcomes, conditional on a tract saturation. For example, the
population average direct causal effect under treatment intensity
\(\psi\), tract size \(n_{tr}\), and \(Tr\) tracts is now defined as
\(\tau^D(\psi\psi) = \frac{1}{Tr}\sum_{t_r = 1}^{Tr}\left[\frac{1}{n_{tr}} \sum_{i = 1}^{n_{tr}} y_{it_r}(1; \psi\psi) - \frac{1}{n_{tr}} \sum_{i = 1}^{n_{tr}} y_{it_r}(0; \psi\psi)\right]\).
The notation \(\tau^D(\psi\psi)\) implies that the realized saturation
for unit \(i\)'s group and the other group in \(i\)'s tract is \(\psi\).
(In this example, \(4/6\) of the units in tract \(t_r\) are assigned to
the treatment condition, given that groups and tracts are of equal
size.) However, when the interference dependency structure is
misspecified, particularly, when partial interference does not hold,
\(\widehat{\tau}^D(\psi)\) is a biased estimator for
\(\tau^D(\psi\psi)\), given that it includes the observed outcomes of
units in groups with realized saturation \(\psi\), in tracts where other
groups are assigned saturation \(\phi\). In other words, the estimator
for the population average potential outcome under saturation \(\psi\)
is
\[\widehat{y}(z; \psi)=\frac{\sum_{g = 1}^{G} \widehat{y}_{g}(z; \psi\psi)\I(S_g = \psi\psi)}{\sum_{g = 1}^{G}\I(S_g = \psi\psi)}  + \frac{\sum_{g = 1}^{G} \widehat{y}_{g}(z; \psi\phi)\I(S_g = \psi\phi)}{\sum_{g = 1}^{G}\I(S_g = \psi\phi)}.\]
The average outcome across groups includes a mixture of the average of
observed outcomes for units with treatment \(z\) in a group with
realized saturation \(\psi\) in a tract in which the realized saturation
for the other group is \(\psi\), and the average of observed outcomes
for units with treatment \(z\) in a group with realized saturation
\(\psi\) in a tract in which the realized saturation for the other group
is \(\phi\).

\begin{figure}

{\centering \includegraphics{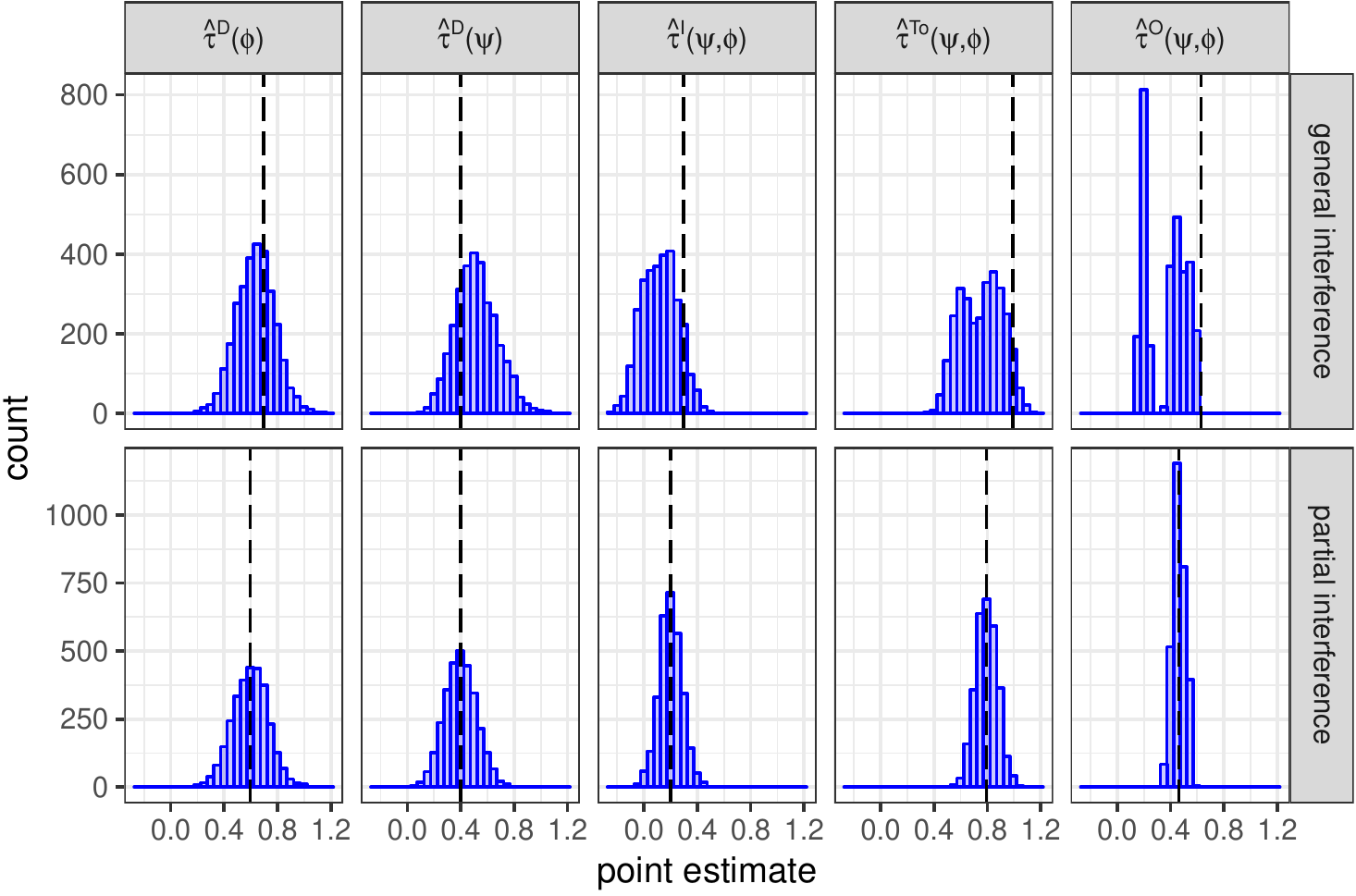} 

}

\caption{\label{fig:miss_hh}The plot shows the distribution of point estimates for 3000 simulations with partial interference specified at the level of groups only. In the upper-row the data generating process is with interference within and across groups (general interference) and in the bottom-row within groups only (partial interference). The dashed vertical line represents the value of the estimand.}\label{fig:unnamed-chunk-17}
\end{figure}

In Figure \ref{fig:miss_hh} we compare the estimators of the direct,
indirect, total and overall causal effects when the partial interference
is misspecified (top row) on the basis of the data generating process
defined in the preceding paragraphs, and when it is correctly specified
(bottom row). The estimators are unbiased when partial interference
holds. However, the estimators are biased when partial interference does
not hold, but the experimenter assumes that it does (top row).

\subsection{Empirical Studies}\label{empirical-studies-1}

\citet{sinclair2012detecting} assess peer effects within a large scale
voter-mobilization hierarchical experiment (of about 70,000 individuals)
conducted in Chicago during a special election in 2009, using as
treatment social-pressure mailings which are sent shortly before the
election and disclose whether a member of the household has voted in
prior elections. The goal is to estimate total causal effects of
mailings and indirect causal effects across and within households. In
the first stage of randomization, neighborhoods (which are about
equal-sized with at most 15 households) are assigned to one of four
different saturations: 100\% of households in the neighborhood are
treated, 50\%, only one, or zero. In the second stage, households are
randomly assigned to treatment according to the neighborhood saturation,
and exactly one individual within each household is randomly selected to
receive the social-pressure message. Therefore, in one-person
households, that person has probability one of being treated, and in
two-person and three-person households, each person has a one-half and
one-third probability of receiving the treatment, respectively. The key
assumption is partial interference at the level of neighborhoods. The
authors find positive total causal effects of receiving the message on
turnout and some evidence of within-household effects but no evidence of
interference across households.

\citet{nickerson2008voting} runs a two-stage placebo-controlled
experiment where in the first stage, households with two registered
voters are assigned to either a 50\% saturation, a 0\% placebo, or a 0\%
pure control saturation. In households assigned to the 50\% saturation,
residents who answer the door receive a face-to-face get-out-the-vote
message (GOTV condition), whereas in households assigned to the 0\%
placebo, residents who answer get a recycling pitch (recycling
condition). The experiment was conducted during the 2002 Congressional
primaries in Denver and Minneapolis. Households were contacted the
weekend before the election. This design identifies the average direct
effect of mobilization with a difference-in-means estimator between
observed voter turnout among reachable voters (those who answer the
door) across the GOTV and recycling conditions. Likewise, the average
indirect effect is the difference-in-means between observed voter
turnout among unreachable residents (those who did not answer the door)
across the GOTV condition and recycling conditions. The author finds a
statistically significant increase in turnout of unreachable residents
in GOTV households of about 6 percentage points, and of 9.8 percentage
points among reachable residents, suggesting that when one voter opens
the door to a canvasser, 60\% of the effect from the get-out-the-vote
appeal is transmitted to the other household member.

With the same design, \citet{bond2018contagion} analyzes the data of a
face-to-face canvassing experiment by \citet{broockman2016durably}
encouraging active perspective taking intended to reduce transphobia.
The effects of the perspective taking exercise spill over to household
residents who do not answer the door, reducing their anti-transgender
prejudice. Other empirical studies using a hierarchical design include
\citet{duflo2003role}, who analyze spillover effects in individuals'
choice of retirement plan, \citet{miguel2004worms} that study the
effects of a deworming medical treatment on health and school
participation of untreated children in (partially) treated schools and
of children in neighboring control schools. Similarly,
\citet{angelucci2009indirect} analyze indirect effects of the cash
transfer program Progresa on consumption. In this case, only the first
stage is randomized assigning municipalities to treatment or control,
while in the second stage a subset of individuals are offered treatment
based on their income. This design identifies direct and indirect
effects when there is partial interference, but cannot identify whether
these effects vary with intensity of treatment because there is no
exogenous variation in treatment saturation. (This is also true of
\citet{nickerson2008voting}'s and \citet{duflo2003role}'s design in
which the group-level saturations are fixed at 50\%.) A similar case is
presented by \citet{sur2009cluster}, who study the effectiveness of a
typhoid vaccine with a design that randomly assigns geographic clusters
to receive the vaccine or a placebo vaccine, but individuals
self-selected into treatment in the second stage. Likewise, the design
by \citet{wilke2019placebo} to study the effectiveness of
education-entertainment on attitudes towards violence against women,
teacher absenteeism, and abortion stigma, first assigned villages to
treatments and in the second stage individuals self-selected into
treatment. \citet{crepon2013labor} implement a two-stage hierarchical
design in the context of a job placement assistance program in France in
which cities are assigned to either one of four positive saturations or
a control condition, and then job seekers are randomly assigned to
treatment according to their city saturation. The program has positive
direct effects, but negative spillover effects: the likelihood of
finding a stable job for untreated job seekers in positive saturation
cities is smaller than for untreated job seekers in control areas.

\citet{bhatti2017voter} combine a hierarchical design with network data
on family ties to assess spillover effects of a GOTV experiment that
mobilized young voters with text messages during municipal and European
Parliament elections in Denmark, and \citet{gine2018together} assess
direct and indirect effects of a voter awareness campaign on female
turnout and candidate choice in Pakistan by assigning first geographical
clusters within villages to either one of two treatments or to control
(with unequal probabilities because the number of clusters varied by
village) and then randomly targeting a subset of households in treatment
clusters. \citet{basse2018analyzing} evaluate an intervention to reduce
student absenteeism in Philadelphia with a two-stage randomization
design in which households with multiple students were first assigned to
treatment or control, then exactly one student in treatment households
was randomly selected for the student's parents to receive
student-specific information. The authors address the practical problem
presented to researchers when household size varies (or when the number
of geographical clusters across villages varies as in
\citet{gine2018together}): whether to assign equal weight to individuals
or to households, depending on what is relevant from a policy
perspective. The authors propose unbiased estimators for individual and
household weighted estimands.

\section{Contagion}\label{contagion}

By analogy to biological contagion, (social) contagion refers to
processes through which the outcomes of one unit causally affect the
outcomes of another unit. Such processes were illustrated above in
Figure \ref{fig:interference_dag} as the path \(Y_1 \rightarrow Y_2\).
In these narrow terms, contagion is distinct from interference. That
said, contagion can be a mechanism through which interference occurs,
and one that may call into question assuming a particular exposure
mapping that limits the extent of interference
\citep{manski2013identification, eckles2017design}. Assessing whether
spillover effects are due to contagion amounts to conducting a mediation
analysis, where the mediators are treated units' outcomes
\citep{vanderweele2012components, ogburn2017vaccines}. This was
displayed in Figure \ref{fig:interference_dag} as the path
\(Z_1 \rightarrow Y_1 \rightarrow Y_2\), where \(Y_1\) is the mediator
in the contagion process. To evaluate whether spillover effects are due
to contagion, conditions must hold so as to identify both spillover
effects (as discussed in this chapter) as well as mediation effects.
Identifying mediation effects requires that other types of conditional
independence hold, such as sequential ignorabilty
\citep{imai-etal2010-mediation, pearl2014}. These are strong assumptions
about the data generating process and typically cannot be induced
directly by an experimental design
\citep{imai-etal2012-mechanisms-experiments}.

\citet{imaiidentification} re-analyze the two-stage placebo-controlled
get-out-the-vote (GOTV) experiment by \citet{nickerson2008voting}. Their
goal is to analyze whether canvassing increases the turnout of the voter
who does not answer via effects on the vote intention of the reachable
voter (contagion), or via other channels such as conversations within
the household (non-contagion spillover). They start with a decomposition
of the indirect effect into the sum of a \emph{contagion
effect}---canvassing influences the turnout of the untreated voter of a
contacted household by changing the vote intention of the treated voter
(approximated by turnout)---and effects due to other mechanisms. With
reference to the setting of \citet{nickerson2008voting}, this
decomposition is based on the following: there is no spillover across
households, reachable voters form a vote intention immediately after
being contacted, where we denote potential vote intention as
\(y_{i1}^*(z_i) \in \{0,1\}\), and the turnout of all voters and all
households is observed. Let \(y_{i1}(z_i)\) and
\(y_{i2}(z_i, y_{i1}^*(z_i))\) represent the potential voting outcome of
reachable and unreachable voters in complier household \(i\),
respectively. Causal quantities of interest are defined only for
compliers (that is, residents of households where someone answers the
door). As such, the analysis is limited to households in which someone
answered the door either in the canvassing or the placebo (recycling)
conditions.

The indirect effect of the GOTV campaign is thus defined as the
difference in the potential outcomes of unreachable voters when their
household is assigned to the GOTV condition (\(z_i=1\)) as opposed to a
control condition (\(z_i=0\)). The average indirect effect is obtained
by taking the average difference across complier households and can be
expressed (in terms of a finite population of \(N_c\) complier
households) as \[
\theta= \frac{1}{N_c}\sum_{i = 1}^{N_c} y_{i2}(1, y_{i1}^*(1)) - y_{i2}(0, y_{i1}^*(0)).
\] Then, \(\theta\) is decomposed into the sum of the contagion effect
and the effect of other mechanisms by considering the vote intention of
the treated voter \(y_{i1}^*\) as the mediator.
\citet{imaiidentification} define an ``average contagion effect'' as
follows (again, written in terms of a finite population of \(N_c\)
complier households): \[
\gamma(z) = \frac{1}{N_c} \sum_{i=1}^{N_c} y_{i2}(z, y_{i1}^*(1)) - y_{i2}(z, y_{i1}^*(0)).
\] Note that it is possible for \(y_{i1}^*(1) = y_{i1}^*(0)\), in which
case there would be no contagion effect for household \(i\). The
quantity \(\gamma(z)\) thus aggregates over cases where there could or
could not be contagion effects, and then for the former, over the
magnitude of any contagion effect. It is therefore analogous to what is
known as a ``natural'' effect, rather than a ``controlled'' effect, in
the mediation literature \citep{imai-etal2010-mediation}. Other,
non-contagion mechanisms are captured by \[
\eta(z)=\frac{1}{N_c} \sum_{i=1}^{N_c} y_{i2}(1, y_{i1}^*(z)) - y_{i2}(0, y_{i1}^*(z)).
\] The indirect effect can be decomposed as
\(\theta = \gamma(1) + \eta(0) = \gamma(0) + \eta(1)\).

Identification of this decomposition requires a sequential ignorability
assumption from mediation analysis. Such assumption implies that
conditional on treatment status and pre-treatment covariates, the vote
intention of the treated voter is independent of the potential outcome
of the unreachable voter. This assumption would be violated in the
presence of unobserved confounders (such as political efficacy) that
affect both the vote intention of the reachable voter and turnout of the
unreachable voter. It is important to note that this assumption is
stronger than the usual ignorability condition necessary for
observational studies because it requires so-called ``cross-world
independence assumptions''---i.e., assumptions about potential outcomes
that can never be revealed by experimentation. Under this key
assumption, \citet{imaiidentification} estimate that indirect effects
can be largely explained by contagion, even for households whose treated
voter is a Democrat and unreachable voter is a Republican. Because this
mediation analysis involves such strong assumptions, the authors
reasonably conduct a sensitivity analysis, examining how robust these
conclusions are to violations of the sequential ignorability assumption.

Other empirical studies exploring contagion effects include
\citet{forastiere2016identification}, who analyze contagion of an
encouragement program on households' use of bed nets in Zambia. Relying
on semi-parametric outcome models, \citet{ferrali2018peer} study the
effects of an encouragement campaign on the adoption of a new political
communication technology in Uganda, and \citet{vasquez2018excombatants}
leverages exogenous shocks to analyze contagion effects of criminal
behavior among ex-combatants in Colombia. One way analysis of contagion
can avoid the strong sequential ignorability assumptions is to instead
assume complete mediation (i.e., an exclusion restriction of
instrumental variables estimation) such that all spillover effects are
due to contagion via a particular outcome; for example,
\citet{eckles2016estimating} conduct an experiment in which they posit
that treatment of an individual's peers only affects them via specific
directed behaviors.

\section{Conclusion}\label{conclusion}

Standard methods for analyzing experiments assume no interference, which
assumes that a unit's own treatment status is all that one needs to know
to characterize its outcome. In many settings, including many of the
empirical examples discussed in this chapter, such an assumption is
unwarranted. Such spillovers may merely represent a nuisance for
estimating quantities of interest. In such cases, experimenters may want
to introduce adjustments to their designs so as to minimize the
potential for exposure to other units. Alternatively, experimenters
could try working at a higher level of aggregation at which interference
is less likely to be a concern. On the other hand, researchers may have
a substantive interest in estimating spillover effects. This chapter
presumed such an interest and proposed methods for doing so.

We reviewed two analytical frameworks for estimating spillover effects
in experiments. In the first, the structure of interference is known but
can be of almost arbitrary form. In the second, the interference
structure is mostly unknown except that the experimenter can be
confident that interference is fully contained within non-overlapping
groups. We demonstrated how one can work under either framework to
estimate spillover effects using the \emph{interference} R package. We
also illustrate the implications of specifying the nature and extent of
interference.

Our review of empirical studies demonstrates the relevance of spillover
effects to various social phenomena, such as voting, petitioning,
student behavior, norms against violence, prejudice, economic decisions,
and subjective well-being. These studies also offer examples of designs
that operationalize the analytical frameworks. We hope that the
analytical foundation and examples provided can help experimenters to
push both the methodological and empirical frontiers in our
understanding of spillover effects.

\renewcommand\refname{References}
\bibliography{references.bib}

@article{imai-etal2012-mechanisms-experiments,
	Author = {Imai, Kosuke and Tingley, Dustin and Yamamoto, Teppei},
	Doi = {10.1111/j.1467-985X.2012.01032.x},
	Eprint = {https://rss.onlinelibrary.wiley.com/doi/pdf/10.1111/j.1467-985X.2012.01032.x},
	Journal = {Journal of the Royal Statistical Society: Series A (Statistics in Society)},
	Number = {1},
	Pages = {5-51},
	Title = {Experimental designs for identifying causal mechanisms},
	Url = {https://rss.onlinelibrary.wiley.com/doi/abs/10.1111/j.1467-985X.2012.01032.x},
	Volume = {176},
	Year = {2013}}

@article{pearl2014,
	Author = {Judea Pearl},
	Journal = {Psychological Methods},
	Number = {4},
	Pages = {459--481},
	Title = {Interpretation and Identification of Causal Mediation},
	Volume = {19},
	Year = {2014}}

@article{imai-etal2010-mediation,
	Author = {Imai, Kosuke and Keele, Luke and Yamamoto, Teppei},
	Journal = {Statistical Science},
	Number = {1},
	Pages = {51--71},
	Title = {Identification, Inference, and Sensitivity Analysis for Causal Mediation Effects},
	Volume = {25},
	Year = {2010}}

@article{taylor-eckles-social-influence-experiments,
	Author = {Sean J. Taylor and Dean Eckles},
	Journal = {arXiv:1709.0963v1 [cs.SI]},
	Title = {Randomized Experiments to Detect and Estimate Social Influence in Networks},
	Year = {2017}}

@article{rosenbaum99-dilated,
	Author = {Paul R. Rosenbaum},
	Journal = {Biometrics},
	Number = {2},
	Pages = {560-564},
	Title = {Reduced sensitivity to hidden bias at upper quantiles in observational studies with dilated treatment effects},
	Volume = {55},
	Year = {1999}}

@misc{zonszein-interference,
	Author = {Stephanie Zonszein and Peter Aronow and Cyrus Samii},
	Title = {{\it interference}, An R Package for Design-Based Estimation of Spillover Effects},
	Year = {2019},
	Note = {Version 0.1.0-alpha},
	version = {0.1.0-alpha},
	url = {https://github.com/szonszein/interference}
	}

@article{liu_hudgens2014,
	Author = {Lan Liu and Michael G. Hudgens},
	Journal = {Journal of the American Statistical Association},
	Number = {505},
	Pages = {288-301},
	Title = {Large Sample Randomization Inference of Causal Effects in the Presence of Interference},
	Volume = {109},
	Year = {2014}}

@article{sandefur2018bloomberg,
	Author = {Justin Sandefur},
	Journal = {Bloomberg Quint Opinion},
	Title = {Cash Transfers Cure Poverty. Side-Effects Vary. Symptoms May Return When Treatment Stops.},
	Volume = {April 22 2018},
	Year = {2018}}

@inproceedings{ugander2013graph,
	Author = {Ugander, Johan and Karrer, Brian and Backstrom, Lars and Kleinberg, Jon},
	Booktitle = {Proceedings of the 19th ACM SIGKDD international conference on Knowledge discovery and data mining},
	Organization = {ACM},
	Pages = {329--337},
	Title = {Graph cluster randomization: Network exposure to multiple universes},
	Year = {2013}}

@inproceedings{saveski2017detecting,
  title={Detecting network effects: Randomizing over randomized experiments},
  author={Saveski, Martin and Pouget-Abadie, Jean and Saint-Jacques, Guillaume and Duan, Weitao and Ghosh, Souvik and Xu, Ya and Airoldi, Edoardo M},
  booktitle={Proceedings of the 23rd ACM SIGKDD International Conference on Knowledge Discovery and Data Mining},
  pages={1027--1035},
  year={2017},
  organization={ACM}
}

@article{tchetgen2012causal,
	Author = {Tchetgen, Eric J Tchetgen and VanderWeele, Tyler J},
	Journal = {Statistical methods in medical research},
	Number = {1},
	Pages = {55--75},
	Publisher = {Sage Publications Sage UK: London, England},
	Title = {On causal inference in the presence of interference},
	Volume = {21},
	Year = {2012}}

@article{liu2014large,
	Author = {Liu, Lan and Hudgens, Michael G},
	Journal = {Journal of the american statistical association},
	Number = {505},
	Pages = {288--301},
	Publisher = {Taylor \& Francis},
	Title = {Large sample randomization inference of causal effects in the presence of interference},
	Volume = {109},
	Year = {2014}}

@article{baird2017optimal,
	Author = {Baird, Sarah and Bohren, J Aislinn and McIntosh, Craig and {\"O}zler, Berk},
	Journal = {Review of Economics and Statistics},
	Number = {0},
	Publisher = {MIT Press},
	Title = {Optimal design of experiments in the presence of interference},
	Year = {2017}}

@article{sinclair2012detecting,
	Author = {Sinclair, Betsy and McConnell, Margaret and Green, Donald P},
	Journal = {American Journal of Political Science},
	Number = {4},
	Pages = {1055--1069},
	Publisher = {Wiley Online Library},
	Title = {Detecting spillover effects: Design and analysis of multilevel experiments},
	Volume = {56},
	Year = {2012}}

@article{bond2018contagion,
	Author = {Bond, Robert M},
	Journal = {Social influence},
	Number = {2},
	Pages = {104--116},
	Publisher = {Taylor \& Francis},
	Title = {Contagion in social attitudes about prejudice},
	Volume = {13},
	Year = {2018}}

@article{sacerdote2001peer,
	Author = {Sacerdote, Bruce},
	Journal = {The Quarterly journal of economics},
	Number = {2},
	Pages = {681--704},
	Publisher = {MIT Press},
	Title = {Peer effects with random assignment: Results for Dartmouth roommates},
	Volume = {116},
	Year = {2001}}

@article{baicker2005spillover,
	Author = {Baicker, Katherine},
	Journal = {Journal of public economics},
	Number = {2-3},
	Pages = {529--544},
	Publisher = {Elsevier},
	Title = {The spillover effects of state spending},
	Volume = {89},
	Year = {2005}}

@article{bond201261,
	Author = {Bond, Robert M and Fariss, Christopher J and Jones, Jason J and Kramer, Adam DI and Marlow, Cameron and Settle, Jaime E and Fowler, James H},
	Journal = {Nature},
	Number = {7415},
	Pages = {295},
	Publisher = {Nature Publishing Group},
	Title = {A 61-million-person experiment in social influence and political mobilization},
	Volume = {489},
	Year = {2012}}

@article{isen2014local,
	Author = {Isen, Adam},
	Journal = {Journal of Public Economics},
	Pages = {57--73},
	Publisher = {Elsevier},
	Title = {Do local government fiscal spillovers exist? Evidence from counties, municipalities, and school districts},
	Volume = {110},
	Year = {2014}}

@article{green2016effects,
	Author = {Green, Donald P and Krasno, Jonathan S and Coppock, Alexander and Farrer, Benjamin D and Lenoir, Brandon and Zingher, Joshua N},
	Journal = {Electoral Studies},
	Pages = {143--150},
	Publisher = {Elsevier},
	Title = {The effects of lawn signs on vote outcomes: Results from four randomized field experiments},
	Volume = {41},
	Year = {2016}}

@article{eckles2016estimating,
	Author = {Eckles, Dean and Kizilcec, Ren{\'e} F and Bakshy, Eytan},
	Journal = {Proceedings of the National Academy of Sciences},
	Number = {27},
	Pages = {7316--7322},
	Publisher = {National Acad Sciences},
	Title = {Estimating peer effects in networks with peer encouragement designs},
	Volume = {113},
	Year = {2016}}

@article{coppock2016treatments,
	Author = {Coppock, Alexander and Guess, Andrew and Ternovski, John},
	Journal = {Political Behavior},
	Number = {1},
	Pages = {105--128},
	Publisher = {Springer},
	Title = {When treatments are tweets: A network mobilization experiment over twitter},
	Volume = {38},
	Year = {2016}}

@article{jones2017social,
	Author = {Jones, Jason J and Bond, Robert M and Bakshy, Eytan and Eckles, Dean and Fowler, James H},
	Journal = {PloS one},
	Number = {4},
	Pages = {e0173851},
	Publisher = {Public Library of Science},
	Title = {Social influence and political mobilization: Further evidence from a randomized experiment in the 2012 US presidential election},
	Volume = {12},
	Year = {2017}}

@inproceedings{kempe2003maximizing,
	Author = {Kempe, David and Kleinberg, Jon and Tardos, {\'E}va},
	Booktitle = {Proceedings of the ninth ACM SIGKDD international conference on Knowledge discovery and data mining},
	Organization = {ACM},
	Pages = {137--146},
	Title = {Maximizing the spread of influence through a social network},
	Year = {2003}}

@article{banerjee2013diffusion,
	Author = {Banerjee, Abhijit and Chandrasekhar, Arun G and Duflo, Esther and Jackson, Matthew O},
	Journal = {Science},
	Number = {6144},
	Pages = {1236498},
	Publisher = {American Association for the Advancement of Science},
	Title = {The diffusion of microfinance},
	Volume = {341},
	Year = {2013}}

@article{banerjee2019using,
	Author = {Banerjee, Abhijit and Chandrasekhar, Arun G and Duflo, Esther and Jackson, Matthew O},
	Journal = {The Review of Economic Studies},
	Title = {Using gossips to spread information: Theory and evidence from two randomized controlled trials},
	Note = {Forthcoming},
	Year = {2019}}

@article{kim2015social,
	Author = {Kim, David A and Hwong, Alison R and Stafford, Derek and Hughes, D Alex and O'Malley, A James and Fowler, James H and Christakis, Nicholas A},
	Journal = {The Lancet},
	Number = {9989},
	Pages = {145--153},
	Publisher = {Elsevier},
	Title = {Social network targeting to maximise population behaviour change: a cluster randomised controlled trial},
	Volume = {386},
	Year = {2015}}

@article{duflo2003role,
	Author = {Duflo, Esther and Saez, Emmanuel},
	Journal = {The Quarterly journal of economics},
	Number = {3},
	Pages = {815--842},
	Publisher = {MIT Press},
	Title = {The role of information and social interactions in retirement plan decisions: Evidence from a randomized experiment},
	Volume = {118},
	Year = {2003}}

@article{miguel2004worms,
	Author = {Miguel, Edward and Kremer, Michael},
	Journal = {Econometrica},
	Number = {1},
	Pages = {159--217},
	Publisher = {Wiley Online Library},
	Title = {Worms: identifying impacts on education and health in the presence of treatment externalities},
	Volume = {72},
	Year = {2004}}

@article{nickerson2008voting,
	Author = {Nickerson, David W},
	Journal = {American political Science review},
	Number = {1},
	Pages = {49--57},
	Publisher = {Cambridge University Press},
	Title = {Is voting contagious? Evidence from two field experiments},
	Volume = {102},
	Year = {2008}}

@article{angelucci2009indirect,
	Author = {Angelucci, Manuela and De Giorgi, Giacomo},
	Journal = {American Economic Review},
	Number = {1},
	Pages = {486--508},
	Title = {Indirect effects of an aid program: how do cash transfers affect ineligibles' consumption?},
	Volume = {99},
	Year = {2009}}

@article{sur2009cluster,
	Author = {Sur, Dipika and Ochiai, R Leon and Bhattacharya, Sujit K and Ganguly, Nirmal K and Ali, Mohammad and Manna, Byomkesh and Dutta, Shanta and Donner, Allan and Kanungo, Suman and Park, Jin Kyung and others},
	Journal = {New England Journal of Medicine},
	Number = {4},
	Pages = {335--344},
	Publisher = {Mass Medical Soc},
	Title = {A cluster-randomized effectiveness trial of Vi typhoid vaccine in India},
	Volume = {361},
	Year = {2009}}

@article{crepon2013labor,
	Author = {Cr{\'e}pon, Bruno and Duflo, Esther and Gurgand, Marc and Rathelot, Roland and Zamora, Philippe},
	Journal = {The Quarterly Journal of Economics},
	Number = {2},
	Pages = {531--580},
	Publisher = {MIT Press},
	Title = {Do labor market policies have displacement effects? Evidence from a clustered randomized experiment},
	Volume = {128},
	Year = {2013}}

@article{forastiere2016identification,
	Author = {Forastiere, Laura and Mealli, Fabrizia and VanderWeele, Tyler J},
	Journal = {Journal of the American Statistical Association},
	Number = {514},
	Pages = {510--525},
	Publisher = {Taylor \& Francis},
	Title = {Identification and estimation of causal mechanisms in clustered encouragement designs: Disentangling bed nets using Bayesian principal stratification},
	Volume = {111},
	Year = {2016}}

@article{bhatti2017voter,
	Author = {Bhatti, Yosef and Dahlgaard, Jens Olav and Hansen, Jonas Hedegaard and Hansen, Kasper M},
	Journal = {Electoral Studies},
	Pages = {39--49},
	Publisher = {Elsevier},
	Title = {How voter mobilization from short text messages travels within households and families: Evidence from two nationwide field experiments},
	Volume = {50},
	Year = {2017}}

@article{basse2018analyzing,
	Author = {Basse, Guillaume and Feller, Avi},
	Journal = {Journal of the American Statistical Association},
	Number = {521},
	Pages = {41--55},
	Publisher = {Taylor \& Francis},
	Title = {Analyzing two-stage experiments in the presence of interference},
	Volume = {113},
	Year = {2018}}

@article{gine2018together,
	Author = {Gin{\'e}, Xavier and Mansuri, Ghazala},
	Journal = {American Economic Journal: Applied Economics},
	Number = {1},
	Pages = {207--35},
	Title = {Together we will: experimental evidence on female voting behavior in Pakistan},
	Volume = {10},
	Year = {2018}}

@article{vanderweele2012components,
	Author = {VanderWeele, Tyler J and Tchetgen, Eric J Tchetgen and Halloran, M Elizabeth},
	Journal = {Epidemiology},
	Number = {5},
	Pages = {751},
	Publisher = {NIH Public Access},
	Title = {Components of the indirect effect in vaccine trials: Identification of contagion and infectiousness effects},
	Volume = {23},
	Year = {2012}}

@article{ogburn2017vaccines,
	Author = {Ogburn, Elizabeth L and VanderWeele, Tyler J},
	Journal = {The Annals of Applied Statistics},
	Number = {2},
	Pages = {919--948},
	Publisher = {Institute of Mathematical Statistics},
	Title = {Vaccines, contagion, and social networks},
	Volume = {11},
	Year = {2017}}

@article{imaiidentification,
	Author = {Imai, Kosuke and Jiang, Zhichao},
	Journal = {Journal of the Royal Statistical Society, Series A},
	Number = {in press},
	Title = {Identification and Sensitivity Analysis of Contagion Effects in Randomized Placebo-Controlled Trials},
	Year = {2019}}

@techreport{beaman2018can,
	Author = {Beaman, Lori and BenYishay, Ariel and Magruder, Jeremy and Mobarak, Ahmed Mushfiq},
	Institution = {National Bureau of Economic Research},
	Title = {Can Network Theory-based Targeting Increase Technology Adoption?},
	Year = {2018},
	Number= {24912}
	}

@article{beaman2018diffusion,
	Author = {Beaman, Lori and Dillon, Andrew},
	Journal = {Journal of Development Economics},
	Pages = {147--161},
	Publisher = {Elsevier},
	Title = {Diffusion of agricultural information within social networks: Evidence on gender inequalities from Mali},
	Volume = {133},
	Year = {2018}}

@article{chin2018evaluating,
	Author = {Chin, Alex and Eckles, Dean and Ugander, Johan},
	Journal = {arXiv preprint arXiv:1809.09561},
	Title = {Evaluating stochastic seeding strategies in networks},
	Year = {2018}}

@article{paluck2016changing,
	Author = {Paluck, Elizabeth Levy and Shepherd, Hana and Aronow, Peter M},
	Journal = {Proceedings of the National Academy of Sciences},
	Number = {3},
	Pages = {566--571},
	Publisher = {National Acad Sciences},
	Title = {Changing climates of conflict: A social network experiment in 56 schools},
	Volume = {113},
	Year = {2016}}

@article{manski2013identification,
	Author = {Manski, Charles F},
	Journal = {The Econometrics Journal},
	Number = {1},
	Pages = {S1--S23},
	Publisher = {Wiley Online Library},
	Title = {Identification of treatment response with social interactions},
	Volume = {16},
	Year = {2013}}

@book{cox58,
	Author = {D. R. Cox},
	Publisher = {Wiley},
	Title = {Planning of Experiments},
	Year = {1958}}

@article{rubin1990,
	Author = {Donald B. Rubin},
	Journal = {Journal of Statistical Planning and Inference},
	Number = {3},
	Pages = {279-292},
	Title = {Formal models of statistical inference for causal effects},
	Volume = {25},
	Year = {1990}}

@article{rosenbaum07_interference,
	Author = {Paul R. Rosenbaum},
	Journal = {Journal of the American Statistical Association},
	Number = {477},
	Pages = {191-200},
	Title = {Interference between units in randomized experiments},
	Volume = {102},
	Year = {2007}}

@article{savje2017average,
	Author = {S{\"a}vje, Fredrik and Aronow, Peter M and Hudgens, Michael G},
	Journal = {arXiv preprint arXiv:1711.06399},
	Title = {Average treatment effects in the presence of unknown interference},
	Year = {2017}}

@article{aronow2012general,
	Author = {Aronow, Peter M},
	Journal = {Sociological Methods \& Research},
	Number = {1},
	Pages = {3--16},
	Publisher = {Sage Publications Sage CA: Los Angeles, CA},
	Title = {A general method for detecting interference between units in randomized experiments},
	Volume = {41},
	Year = {2012}}

@article{bowers2013reasoning,
	Author = {Bowers, Jake and Fredrickson, Mark M and Panagopoulos, Costas},
	Journal = {Political Analysis},
	Number = {1},
	Pages = {97--124},
	Publisher = {Cambridge University Press},
	Title = {Reasoning about interference between units: A general framework},
	Volume = {21},
	Year = {2013}}

@article{athey2018exact,
	Author = {Athey, Susan and Eckles, Dean and Imbens, Guido W},
	Journal = {Journal of the American Statistical Association},
	Number = {521},
	Pages = {230--240},
	Publisher = {Taylor \& Francis},
	Title = {Exact p-values for network interference},
	Volume = {113},
	Year = {2018}}

@article{aronow2017estimating,
	Author = {Aronow, Peter M and Samii, Cyrus},
	Journal = {The Annals of Applied Statistics},
	Number = {4},
	Pages = {1912--1947},
	Publisher = {Institute of Mathematical Statistics},
	Title = {Estimating average causal effects under general interference, with application to a social network experiment},
	Volume = {11},
	Year = {2017}}

@article{eckles2017design,
	Author = {Eckles, Dean and Karrer, Brian and Ugander, Johan},
	Journal = {Journal of Causal Inference},
	Number = {1},
	Publisher = {De Gruyter},
	Title = {Design and analysis of experiments in networks: Reducing bias from interference},
	Volume = {5},
	Year = {2017}}

@article{hudgens2008toward,
	Author = {Hudgens, Michael G and Halloran, M Elizabeth},
	Journal = {Journal of the American Statistical Association},
	Number = {482},
	Pages = {832--842},
	Publisher = {Taylor \& Francis},
	Title = {Toward causal inference with interference},
	Volume = {103},
	Year = {2008}}

@article{haushofer2018long,
	Author = {Haushofer, Johannes and Shapiro, Jeremy},
	Journal = {Busara Center for Behavioral Economics, Nairobi, Kenya},
	Title = {The long-term impact of unconditional cash transfers: Experimental evidence from Kenya},
	Year = {2018}}

@article{ferrali2018peer,
	Author = {Ferrali, Romain and Grossman, Guy and Platas, Melina and Rodden, Jonathan},
	Title = {Peer effects and externalities in technology adoption: Evidence from community reporting in Uganda},
	Year = {2018}}

@article{vasquez2018excombatants,
	Author = {V\'asquez-Cort\'es, Mateo},
	Title = {Criminality as a Social Process: Evidence From Colombian Ex-combatants},
	Year = {2018}}

@article{aronow2013dissertation,
	Author = {Aronow, Peter M},
	Journal = {Dissertation, Yale University, New Haven, CT},
	Title = {Model assisted causal inference},
	Year = {2013}}

@article{egami2017unbiased,
	Author = {Egami, Naoki},
	Journal = {arXiv preprint arXiv:1708.08171},
	Title = {Unbiased Estimation and Sensitivity Analysis for Network-Specific Spillover Effects: Application to An Online Network Experiment},
	Year = {2017}}

@article{rogowski2012estimating,
	Author = {Rogowski, Jon C and Sinclair, Betsy},
	Journal = {Political Analysis},
	Number = {3},
	Pages = {316--328},
	Publisher = {Cambridge University Press},
	Title = {Estimating the causal effects of social interaction with endogenous networks},
	Volume = {20},
	Year = {2012}}

@article{broockman2016durably,
	Author = {Broockman, David and Kalla, Joshua},
	Journal = {Science},
	Number = {6282},
	Pages = {220--224},
	Publisher = {American Association for the Advancement of Science},
	Title = {Durably reducing transphobia: A field experiment on door-to-door canvassing},
	Volume = {352},
	Year = {2016}}

@article{wilke2019placebo,
	Author = {Wilke, Anna and Green, Donald P and Cooper, Jasper},
	Title = {A Placebo Design to Detect Spillovers from an Education-Entertainment Experiment in Uganda},
	Year = {2019}}

@article{sobel2006,
	Author = {Michael E. Sobel},
	Issn = {01621459},
	Journal = {Journal of the American Statistical Association},
	Number = {476},
	Pages = {1398--1407},
	Publisher = {[American Statistical Association, Taylor & Francis, Ltd.]},
	Title = {What Do Randomized Studies of Housing Mobility Demonstrate?: Causal Inference in the Face of Interference},
	Url = {http://www.jstor.org/stable/27639760},
	Volume = {101},
	Year = {2006}}

\end{document}